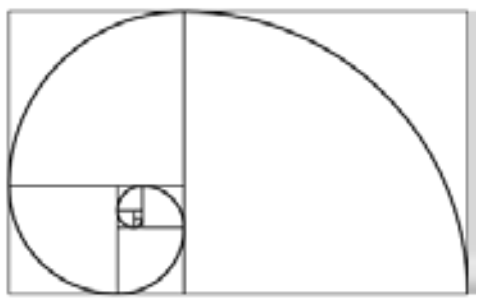

# SPIRALE

A One-Time-Pad Cipher

Philippe Allard[1]
AMD-crypto@orange.fr

## Content

## 1. Introduction

Despite our epoch of high technology and the powerfulness of modern personal computers allowing implementation of highly sophisticated digital ciphers with symmetric or public key, there is still some effort to produce a modern hand cipher for paper&pencil encryption, too. The well-known Solitaire by B. Schneier [2]and the Handycipher by B. Kallick [1] are the most recent ones (historic ones are ADFGVX or double-column transposition).

Solitaire is a one-time pad (OTP) cipher, the modern version of a Vigenère cipher characterized by a random key of same length as plaintext. The proven supremacy of the OTP approach is based upon an absolute randomness of the keystreams and a unique usage. The first condition can be achieved only by using a physical process, like nuclear disintegration, and the second one by physical transmission of keystreams in advance via an absolutely secure channel between the correspondents storing them securely. These conditions are only exceptionally realized, and the practical, and so degraded, implementation of the concept is made by calculating each time a pseudo-random keystream from secret keys shared by the corresponding parties. Solitaire is original by its use of a deck of cards to manually realize the computational process of keystream generating. It is well studied and its security proven.

[1] The author owes deep thanks to Prof. Bernhard Esslinger and his students for their friendly and accurate re-readings and clarifications of this manuscript.

As Solitaire now is well known keeping a deck of cards may be almost as compromising as a personal computer with ciphering programs. So, B. Kallick proposed Handycipher as an alternative solution needing no more than pen and paper. It is a non-deterministic homophonic substitution cipher in which each letter of plaintext is replaced pseudo-randomly by a group of 1 to 5 characters. So the ciphertext is much larger than the plaintext in a rate generally higher than 4. This practical handicap plus to the rather tedious encryption process let us assume that this algorithm is not really satisfying as an alternative.

Thus we propose here another solution. It is also an OTP cipher designed to be simple to implement by hand, with a high level of variability in keys equivalent to a 128-bit key cryptosystem. To generate the pseudo random keystream it is based upon classical features like Fibonacci sequence and congruence but these concepts are revisited to reach the desired level of strength facing cryptanalysis. The final algorithm needs almost no mental calculations and only repetitive entering in a special table. The process is also resilient to errors as they have only a local effect without obscuring all the ciphertext.

Spirale cipher is running through 4 different steps according the following flow chart (input data is blue, fixed algorithms is black, intermediate data is green, and the final result is red):

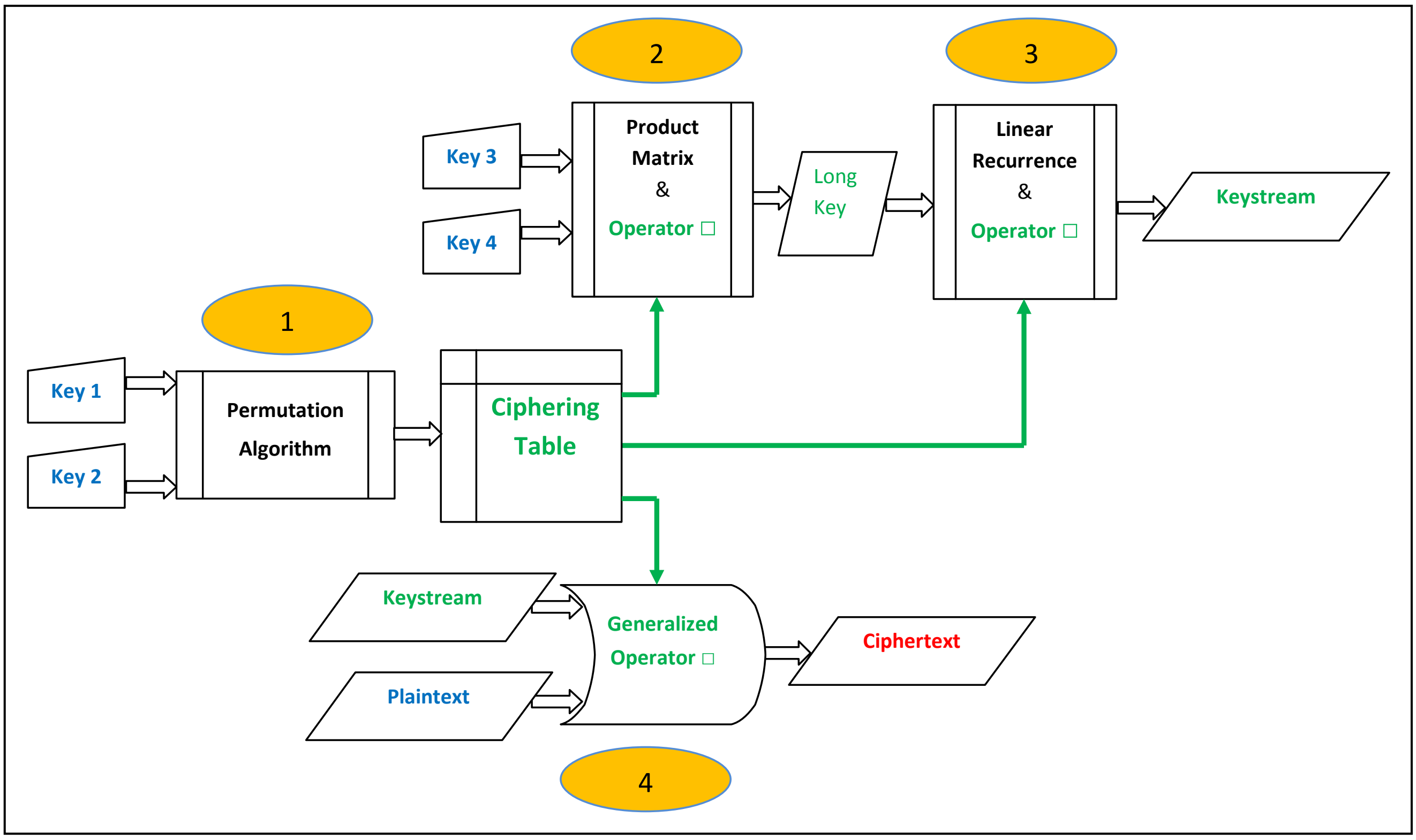

Figure 1

For the naming of the following headlines we used an epistemological choice: The name describes the concepts from which we leave, generalize them along the text and express the synthesis of the new concepts used in Figure 1.

## 2. Generalizing the Fibonacci Sequence (used in algorithm step 3)

At the beginning we work with the classical alphabet A, B, C … X,Y, Z and we associate each letter with its alphabet rank as integer value: 1 to A, 2 to B, …,26 to Z.As Spirale is an OTP cipher its goal is to create a "good" pad of letters. Our first idea is to start from a given string of length k used as a seed key. Working on the letter's integer values means that Spirale creates an arbitrary-length sequence of integers (below in green color) from a given integer sequence of length k (below in blue). The challenge is to generate by a deterministic computation a long sequence of letters as pseudo random as possible and in an easy-to-apply manner. The generation of such a long key will be described in the following chapter3.

Working on the letters' values and fulfilling the requirement to always stay in the interval [1, 26] the calculations must be done with congruence to modulus 26 (arithmetically $26 \equiv 0$).

A classical solution to generate a sequence is to use a recurrence relation [3]. Keeping in mind the goal of an easy to compute hand cipher, the simplest form to apply is the linear recurrence of which the archetype case is the Fibonacci sequence:

$$X_n = (X_{n-1} + X_{n-2}) \bmod 26$$

This results in a number sequence like **16, 19**,9, 2, 11, 13, 24, 11, 9, 20, 3, 23, 26, 23, 23, 20, 17, 11, 2, …(with k=2)which seems random – however, by this formula each element is strongly correlated with the two precedent ones. An option to avoid this feature is to increase the order by taking into account more terms but this increases the burden of calculation. A compromise is to space the only two terms used in recurrence, for example the second one at the opposite side of the initial sequence of length k:

$$X_n = (X_{n-1} + X_{n-k}) \bmod 26$$

With a given integer sequence {16, 19, 9, 2} of length k=4, the first integers are **16, 19, 9, 2**, 18, 11, 20, 22, 14, 25, 19, 15, 3, 2, 21, 10, 13, 15, 10, 20, 7, 22, 6, 26, 6, 2, 8, 8, 14, 16, … Using the above formula means that two consecutive terms are still correlated, as $X_n$ depends on $X_{n-1}$. So, an error in calculating of $X_{n-1}$ (in the example sequence above the last green value 26 is intentionally wrong) propagates to $X_n$ and all following numbers (like in the following six numbers in red).

To avoid this disadvantage we could add the distance k in the index not only to the second operand but also to the first one:

$$X_n = (X_{n-k+1} + X_{n-k}) \bmod 26$$

Or more general, we introduce depth d(d is close to the middle of the initial sequence distance k),to share spatially with uniformity the local influence of the long key's elements on those of the keystream:

$$X_n = (X_{n-d} + X_{n-k}) \bmod 26 \qquad (1)$$

Figure 2 explicates the idea: Number $X_n$ is influenced only by numbers $X_{n-k}$ and $X_{n-d}$: As closer d is to k (for instanced=k-1) as bigger is the contagious influence on consecutive keystream numbers. So, by spacing n-d from n-k we add some more pseudo randomness into the keystream. We also prefer this last solution because so, two consecutive elements are correlated with two pairs of preceding elements completely different and an error will propagate only by jumps of d and k elements. However, one case has to be avoided:

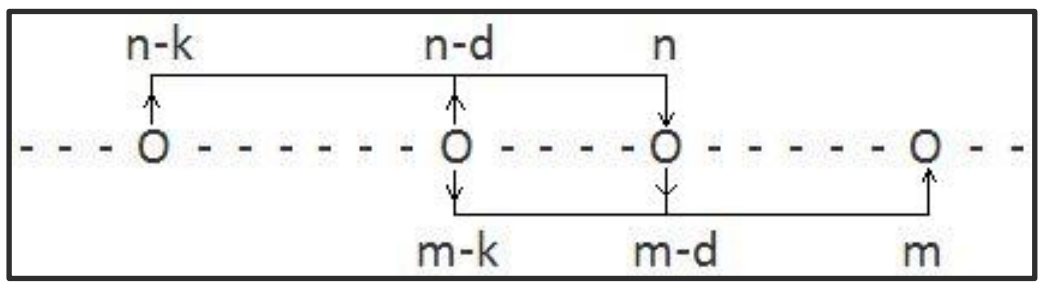

Figure 2

In this case an element m is built from two generated elements which are already dependent. In such a situation we don't get a single and long linear recurrence sequence but a short series of interlaced classical Fibonacci series. The conditions to be avoided for this situation are:

$$n-d = m-k \quad \text{and} \quad n = m-d$$

thus

$$k = 2d \qquad (2)$$

## 3. Long Key Generation (used in algorithm step 2)

To decrease correlation between consecutive elements, we saw that it is better to have a long initial sequence. To create such a long sequence that takes the role of the key in the ciphering process, we use two shorter keys K1 and K2 and organize data and results in a matrix via the following formula:

$$Y_{p,q} = (X_p + X_q) \bmod 26$$

For example with the 4-element K1 = {19, 15, 12, 9} and the 5-element K2 = {20, 1, 9, 18, 5} we generate a 4x5 matrix of 20 cells:

| | | | | | | |
|---|---|---|---|---|---|---|
| Key 1 | 19 | 13 | 20 | 2 | 11 | 24 |
| | 15 | 9 | 16 | 24 | 7 | 20 |
| | 12 | 6 | 13 | 21 | 4 | 17 |
| | 9 | 3 | 10 | 18 | 1 | 14 |
| | | 20 | 1 | 9 | 18 | 5 |
| | | Key 2 | | | | |

Figure 3

Then we could read this matrix horizontally or vertically. However, if one of the keys contains a repeated number this implies a repeated row or column. To avoid this, a first solution is to have no number twice in a key. Another solution is to read in the matrix differently. We decide for reading diagonally from top left corner and ascending. This is one of the classic ways in transposition ciphers to scramble the order of elements in a message. Any other easy and efficient path would be acceptable too. So the generated final long key has k=20elements:

{13,9, 20, 6, 16, 2, 3, 13, 24, 11, 10, 21, 7, 24, 18, 4, 20, 1, 17, 14}

and could be used as initial sequence with the linear recurrence formula:

$$X_n = (X_{n-9} + X_{n-20}) \bmod 26$$

with n-9 to avoid condition (2). So the first generated elements would be:

13,21→ 7    9,7 →16    20,24 → 18    6,18 → 24    …

To slide along the generated sequence and identify the elements to use in formula, it is very useful to make a simple tape of paper with 3 marks, spaced here respectively by 11 and 9 cells – as shown in Figure 4:

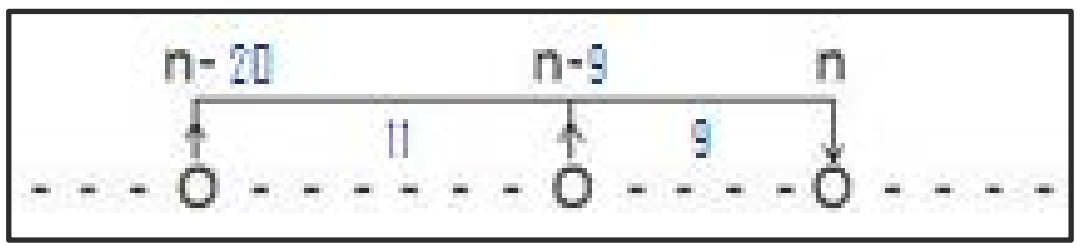

Figure 4

## 4. Generalizing Congruent Addition (used in algorithm steps 2 to 4)

Enciphering of stream ciphers normally uses the XOR operand or a congruent addition.

a) To introduce the idea of a “better” suited operand we first limit us to integers between 0 and 9. The orthodox congruence modulo 10 composed with addition defines a mapping of [0, 9]x[0, 9] to the interval [0, 9] and an internal operator, that we will note ⊕:

$$X \oplus Y = (X+Y) \bmod 10$$

This operator can also be described (pre-calculated) by a mapping matrix:

| ⊕ | 0 | 1 | 2 | 3 | 4 | 5 | 6 | 7 | 8 | 9 |
|---|---|---|---|---|---|---|---|---|---|---|
| 0 | 0 | 1 | 2 | 3 | 4 | 5 | 6 | 7 | 8 | 9 |
| 1 | 1 | 2 | 3 | 4 | 5 | 6 | 7 | 8 | 9 | 0 |
| 2 | 2 | 3 | 4 | 5 | 6 | 7 | 8 | 9 | 0 | 1 |
| 3 | 3 | 4 | 5 | 6 | 7 | 8 | 9 | 0 | 1 | 2 |
| 4 | 4 | 5 | 6 | 7 | 8 | 9 | 0 | 1 | 2 | 3 |
| 5 | 5 | 6 | 7 | 8 | 9 | 0 | 1 | 2 | 3 | 4 |
| 6 | 6 | 7 | 8 | 9 | 0 | 1 | 2 | 3 | 4 | 5 |
| 7 | 7 | 8 | 9 | 0 | 1 | 2 | 3 | 4 | 5 | 6 |
| 8 | 8 | 9 | 0 | 1 | 2 | 3 | 4 | 5 | 6 | 7 |
| 9 | 9 | 0 | 1 | 2 | 3 | 4 | 5 | 6 | 7 | 8 |

Table 1: Mapping matrix with ⊕

It has special properties:

1. Symmetry $X \oplus Y = Y \oplus X$
2. Multiple input pairs (X,Y) give the same result $X \oplus Y$,

   for instance: $3 = 1 \oplus 2 = 5 \oplus 8 = 6 \oplus 7$ $5 = 2 \oplus 3 = 9 \oplus 6 = 7 \oplus 8 \ldots$
3. Each row and column within the table contains all different result values 0, 1, 2, … 9 exactly once; and for each one there is the same number of input pairs creating it.

The last property, a uniform surjectivity, is useful in cryptography, as allowing to build functions hard to inverse or to uncover because of the multiplicity of possible solutions (I guess this statement is evident).For example, if one has the result value 3 in a ciphertext how to guess if it is coming from pair (1,2) or pair (5,8) or pair (6,7) ?

What occurs if we permute the result values inside the matrix? Symmetry will probably be lost (which finally is an advantage) but not the main property: Equal number of result values and equal number of preimages for each result. So we could create another useful function for ciphering. As this matrix has 10x10=100 elements, there are $100! = 9.3 \times 10^{157}$combinations.

This monstrous number is much beyond our needs and we can easily build a sufficient number of efficient functions, now noted □, by simply permuting input values (the indices of the rows and the columns). The use of the two symbols ⊕ and □ is fundamental to show the difference between classical congruent addition and the generalized operation with the permutation of matrix indices.

Here is an example for such a permuted matrix:

| □ | 6 | 9 | 2 | 1 | 8 | 5 | 0 | 4 | 3 | 7 |
|---|---|---|---|---|---|---|---|---|---|---|
| 3 | 0 | 1 | 2 | 3 | 4 | 5 | 6 | 7 | 8 | 9 |
| 2 | 1 | 2 | 3 | 4 | 5 | 6 | 7 | 8 | 9 | 0 |
| 9 | 2 | 3 | 4 | 5 | 6 | 7 | 8 | 9 | 0 | 1 |
| 6 | 3 | 4 | 5 | 6 | 7 | 8 | 9 | 0 | 1 | 2 |
| 1 | 4 | 5 | 6 | 7 | 8 | 9 | 0 | 1 | 2 | 3 |
| 8 | 5 | 6 | 7 | 8 | 9 | 0 | 1 | 2 | 3 | 4 |
| 7 | 6 | 7 | 8 | 9 | 0 | 1 | 2 | 3 | 4 | 5 |
| 5 | 7 | 8 | 9 | 0 | 1 | 2 | 3 | 4 | 5 | 6 |
| 0 | 8 | 9 | 0 | 1 | 2 | 3 | 4 | 5 | 6 | 7 |
| 4 | 9 | 0 | 1 | 2 | 3 | 4 | 5 | 6 | 7 | 8 |

Table 2: Mapping matrix with □

We can see that with this new function:

$3 \square 6 = 0$ and $6 \square 3 = 1$ $1 \square 7 = 3$ and $7 \square 1 = 6$ …

$2 = 3 \square 2 = 2 \square 9 = 9 \square 6 = 1 \square 3 = 8 \square 4 = 7 \square 0 = 5 \square 5$

To read easily from the table it would be useful to re-order the entries by permuting rows and columns like in classical transposition ciphers. We will see later that we can avoid this step.

Permuting independently the entries produces 10!x10! combinations, so more than $1.3x10^{13}$ combinations.

b) Now, that we know how to build a lot of ciphering functions, we can return to letters, alphabet and congruence modulo 26. With letters, the orthodox congruent addition modulo 26 is described by the classic Vigenère table [Appendix A]. By permuting each alphabet entry we can create a completely different *ciphering table*. An example is in Appendix B. To facilitate entries in the table it is not necessary to re-order row and column entries but to add a row and a column filled with the rank of the letters in each permuted alphabet.

The ciphering table is a *look-up table*. That's why Spirale needs no calculations to be applied.

With the sample ciphering table of Table 2, let us encrypt SPIRALE with SGKKFPW:

| | | | | | | | |
|---|---|---|---|---|---|---|---|
| plaintext | S | P | I | R | A | L | E |
| corresponding row | 15 | 16 | 11 | 21 | 13 | 19 | 17 |
| keystream | S | G | K | K | F | P | W |
| corresponding col | 11 | 12 | 2 | 2 | 3 | 19 | 5 |
| letter at intersection | Y | A | L | V | O | K | U |

We observe that the encryption process needs no mental calculations at all, only a series of look-ups in the ciphering table, so it is easy and allows a better performance. Building the ciphering table by alphabet permutations is the first step in the complete ciphering process. This table must be created and used only to encipher or decipher one message. Afterwards it has to be destroyed in order to be applied in the context of secret communications.

The decryption process is inverting the ciphering process: entering from column, reading the right letter of ciphertext and leaving by the same row giving the plaintext letter.

| | | | | | | | |
|---|---|---|---|---|---|---|---|
| keystream | S | G | K | K | F | P | W |
| corresponding col | 11 | 12 | 2 | 2 | 3 | 19 | 5 |
| ciphertext | Y | A | L | V | O | K | U |
| corresponding row | 15 | 16 | 11 | 21 | 13 | 19 | 17 |
| plaintext | S | P | I | R | A | L | E |

From now onwards, we always use a ciphering table (look-up table), derived by a permutation of the Vigenère table and the associated operator □ for generating the long key, the keystream, and the ciphertext (Instead of the simple congruent addition ⊕ used in the introduction at the beginning of chapter 4, the operator □ delivers the result of two table entries. So the matrix in Figure 3 works like a "*product matrix*" of keys generating the long key. An example of such a product matrix is in appendix D3). So, finally the ciphering table is the "kernel" of Spirale.

## 5. Permutation Algorithm (used in algorithm step 1)

We need to permute the alphabet two times per message to send. For this we apply the following algorithm:

- From a short string playing the role of key;
- We derive a *permutation list* of integers filled with the original rank in alphabet of each letter;
- Then the permutation is made by wrapping around the alphabet, beginning at top right end and picking letters spaced according to the permutation list. At left end return to right end. When a letter is already picked, it is jumped at next passages. The permutation list is read cyclically until exhaustion of the original alphabet. The picked letters are progressively tidied up to form the permuted alphabet.

Appendix C describes an example with key BHMAY and its permutation list [2, 8, 13, 1, 25]. Let us examine the process of permutation of the first letters from the original alphabet: 2→Y, 8→Q, 13→D, 1→C, 25→Z by turning around the alphabet until the reading origin, then new reading of permutation list 2→W by jumping Y already picked and crossed, 8→N by jumping Q, 13→V by jumping D and C, 1→U, 25→K by jumping Q-N-D-C-Z-Y-W-V-U, … so the permuted alphabet begins with YQDCZWNVUK… And the array is progressively filled with the new ranks of letters to be used in the ciphering table.

In this step, again, there is no calculation needed, and only a simple and repetitive operation.

## 6. Spirale Cipher (used in algorithm step 4)

We now have all the concepts and tools useful to build our cipher. It needs 4 keys:

- K1 to permute the rows in the ciphering table (see step 1)
- K2 to permute the columns in the ciphering table (see step 1)
- K3 to create rows in the long key matrix (see step 2)
- K4 to create columns in the long key matrix (see step 2)

A key of p letters offers $C=26^p$ combinations. The effects of the four keys are independently combined and interlaced in the ciphering process. The total number of combinations is thus $C_1xC_2xC_3xC_4 = 26^{p_1+p_2+p_3+p_4}$.So the total number of combinations depends only on the cumulated size *n* of these keys formed of sequences of letters ("Cumulating" as in probabilities, independence of causes implies the product of combinations: The permutation of rows (by K1) is made independently of the permutation of columns (by K2); then one can write any string on rows side (K3) of the matrix giving the long key and one can write any string on columns side (K4) of this matrix). The project is to give to Spirale a security level against brute-force attacks equivalent to a 128-bit cipher, and as

$$2^{128} = 26^n \quad \text{implies} \quad n \approx 27.2$$

That means that with a global key size of 28 letters this goal is clearly reached – assuming we have a real random structure for each key and assuming that for each message a new set of keys is used. We decide to uniformly share the combinatory security among all the keys, thus each key must have 7 letters. All the steps already described will be executed with such keys.

Long key generation (in step 2) will be done according formula(3) (X is now a symbol for letters):

$$Y_{p,q} = X_p \square X_q \tag{3}$$

And thus the long key will be 7x7=49 letters long. For a message shorter than 50 letters, this long key will already be the OTP. For a longer message, a keystream has to be generated (in step 3) according formula (4):

$$X_n = X_{n-49} \square X_{n-24} \tag{4}$$

The order of terms is important now as this operation is no more symmetrical.

All that remains is the problem of furnishing securely enough keys to the participating correspondents. We suggest the following procedure to create them in order to avoid dictionary attacks: The corresponding parties select one book, and one page is used per message. In this page they take the 7 first letters and the 7 last letters of the first line and of the last line – whatever the words or punctuation are (many other rules of choice would be suitable too). Then they write them, one extract per row, in a 7x4 array and read it vertically from top right corner. This is equivalent to reverse each string and interlace them uniformly. And finally, they cut the resulting string in four segments of 7 letters.

Example: Text in page: "***We got in***to Milan … unloa***ded us in***[2]

……………

***said th***is had … around ***his neck***."[3]

| | | | | | | |
|---|---|---|---|---|---|---|
| w | e | g | o | t | i | n |
| d | e | d | u | s | i | n |
| s | a | i | d | t | h | i |
| h | i | s | n | e | c | k |

The 4 created keys: nnikiih ctsteou dngdise eaiwdsh

This procedure is quite simple but has a default – it uses a natural language source like a book and not a randomly generated text: The occurrence probability of the letters in keys is so close to that of the book's idiom and thus these probabilities are not equal. A simple manner to partly correct this default could be to replace, in the selected strings, some highest-frequency letters (e,t,a,o,i,n in English) by lowest-frequency letters (z,q,x,j,k,v in English).

[2] The bold parts in this line make up row 1 and row 2.

[3] *A Farewell to Arms*, E. Hemingway, p81, Charles Scribner's Sons edition.

The corrected keys of the above example could be thus: NVIKKIH, CTSQEOU, DNGDKSZ and EAIWDSH.

Now, one can also understand the name given to the cipher: Starting from 4 keys playing the role of seeds, inducing a running around process to build the permuted alphabets and the derived ciphering table that is the core of its security, and finally deploying a potentially infinite keystream. The 4-center spiral (the image at top left corner of the first page of this paper) is just a symbol for this.

## 7. Implementation and Working Example

Several form sheets have been designed for an easy application of Spirale. They are collected in Appendix D and illustrated by a complete working example from the keys above:

- Appendix D1: Alphabet Permutations (step 1)
  To facilitate perception of still eligible letters in the original alphabet we suggest crossing the already permuted letters with a dark pen. And to follow the advancement in cyclical running along the permutation key we suggest putting a mark, a point for example, under a letter each time it is used.

- Appendix D2: Ciphering Table (step 1)
  Just copy in the results from Appendix D1.

- Appendix D3: Long Key Generation& Keystream Generation (step 2 and 3)
  This sheet is designed to generate a keystream up to 400 letters. If you have a longer text just use a second or third copy of this sheet.
  If the long key is too large to apply the precedent suggestion to use a marked tape of paper sliding along the keystream, we arrange the letters so that the cells implicated in recurrence (of ranks n-49 and n-24) are in the same column and can easily be read. The recurrence relation (4) can also be written as

$$X_p \square X_{p+25} = X_{p+49} \qquad (4a)$$

  This implies that lines in the form sheet are 25-cell long and that the result is not in the same column but on the left:

| | | |
|---|---|---|
| … | $X_p$ | … |
| … | $X_{p+25}$ | … |
| $X_{p+49}$ | $X_{p+50}$ | … |

  Into this form sheet we write the results from the top left cells. These shaded cells contain the value missing in the last cell of respective above lines and thus we copy there these values.

- Appendix D4: Encryption or Decryption Process (step 4)
  This sheet is designed to process 200 letters, for a longer text use another sheet. As we use a basic alphabet, the plaintext cannot contain numerals and we have also to remove all spaces and punctuation. The process is executed line by line, top-down for encryption by looking up in ciphering table for respective letters of plaintext and keystream, or bottom-up for decryption by looking up in ciphering table (in the same column) for respective letters of keystream and ciphertext.

Blank versions of these standard forms are collected in Appendix E.
Supported by these sheets, an absolute beginner can work more quickly. So it takes about one hour to perform all the steps and encrypt the 75-letter plaintext example.[4]

## 8. Challenges

To help cryptanalysts in breaking this cipher here we propose four different ciphertexts[5] corresponding to weakened situations of usage:

- Ciphertext 1:
  made from a 314-letter text beginning with the plaintext of the working example.
- Ciphertext 2
  from a 659-letter plaintext encrypted with the same 4 keys as ciphertext 1.
- Ciphertext 3:
  from a 949-letter plaintext encrypted with 2 equal 2 keys (K1 and K2) as in ciphertexts 1 and 2.
- Ciphertext 4:
  from a 485-letter plaintext encrypted with 4 new random keys. All these ciphertexts are in Appendix F.

## 9. Alphabet Extensions

With an alphabet limited to letters, the expression of numbers or dates must be done literally (2015 = two zero one five or two thousands fifteen) that is taking a lot of place in the message. A simple solution to this drawback is to extend the alphabet, for example like:

A … Z 0 1 2 3 4 5 6 7 8 9

This implies no complication in the algorithm: The only difference is to extend the alphabet array in Appendix E1from 26 to36 cells and accordingly in the ciphering table. The long key and the keystream will then be a sequence of letters and numerals. The four 7-character keys could then include numerals too. By this way it is also an increase of combinations for these keys to $36^7$ and also for the permuted alphabets to $36^7 x 36^7$, thus a huge increase in combinatorial security of the cipher.[6]

[4] Also a Python program for decryption and encryption with Spirale is freely available. Its source code is structured according the four steps in Figure 1.
[5]To use Spirale in an operational mind, ciphertexts 2 to 4 are taken from open military intelligence.

The generalization can go further and include also typographic characters to allow readability of message or treatment of peculiar segments like mathematical or chemical formulas and economic data. Such alphabet could be (here character _ is for *space*):

A … Z 0 … 9 _ , . ( ) + - * / ^ < = > % € £ $

In this case, the permutations would be executed on 53 characters and thus extending also the ciphering table and the strength of the cipher against brute-force attacks. The according versions of appendices E1 and E2 for 53 characters are in appendices G1 and G2. Appendix G3 shows an example plaintext using a richer alphabet (59 characters allowing mathematical and chemical formulas) and its ciphertext, produced with an Iron Python program.

As Spirale is designed to be implemented by hand, it could work without any difficulty with any alphabet, even exotic ones:

Α Β Γ Δ Ε Ζ Η Θ Ι Κ Λ Μ Ν Ξ Ο Π Ρ Σ Τ Υ Φ Χ Ψ Ω

ا ب ج د ه و ز ح ط ي ك ل م ن س ع ف ص ق ر ش ت ث خ ذ ض ظ غ

अ आ इ ई उ ऊ ऋ ॠ ए ऐ ओ औ अं अँ अः ऌ ॡ प पा पि पी पु पू पृ पॄ पे पै पो पौ पं पाँ पः पॢ पॣ

It is just necessary to create the corresponding Vigenère table and associate two permuted alphabets to build a ciphering table. For right-to-left writings the form sheets are still useful and only Appendix E3 has to be mirrored to give Appendix E3a.

[6] Together with the Python source code sample alphabet files are delivered: The default alphabet file with 26 letters is alpha_L.txt. The alphabet file with 36 characters is alpha_LN.txt. An alphabet file with 59 printable characters is alpha_LNT.txt.

## 11. Appendices

| | |
|---|---|
| A | Vigenère Table |
| B | Ciphering Table: Example |
| C | Permutation Algorithm: Example |
| D1 | Worked example Alphabet Permutations |
| D2 | Worked example Ciphering Table |
| D3 | Worked example Long Key Generating --- Keystream Generating |
| D4 | Worked example Encryption or Decryption |
| E1 | Alphabet Permutations |
| E2 | Ciphering Table |
| E3 | Long Key Generating --- Keystream Generating |
| E3a | Long Key Generating --- Keystream Generating (for right-to-left writings) |
| E4 | Encryption or Decryption |
| F | Challenges |
| G1 | Extended Alphabet Permutations |
| G2 | Ciphering Table |
| G3 | Example of Application with Extended Alphabet |
| G3a | Interactive Window with the Output of the Python Program |

Appendix A

# Vigenère Table

| ⊕ | A | B | C | D | E | F | G | H | I | J | K | L | M | N | O | P | Q | R | S | T | U | V | W | X | Y | Z |
|---|---|---|---|---|---|---|---|---|---|---|---|---|---|---|---|---|---|---|---|---|---|---|---|---|---|---|
| **A** | A | B | C | D | E | F | G | H | I | J | K | L | M | N | O | P | Q | R | S | T | U | V | W | X | Y | Z |
| **B** | B | C | D | E | F | G | H | I | J | K | L | M | N | O | P | Q | R | S | T | U | V | W | X | Y | Z | A |
| **C** | C | D | E | F | G | H | I | J | K | L | M | N | O | P | Q | R | S | T | U | V | W | X | Y | Z | A | B |
| **D** | D | E | F | G | H | I | J | K | L | M | N | O | P | Q | R | S | T | U | V | W | X | Y | Z | A | B | C |
| **E** | E | F | G | H | I | J | K | L | M | N | O | P | Q | R | S | T | U | V | W | X | Y | Z | A | B | C | D |
| **F** | F | G | H | I | J | K | L | M | N | O | P | Q | R | S | T | U | V | W | X | Y | Z | A | B | C | D | E |
| **G** | G | H | I | J | K | L | M | N | O | P | Q | R | S | T | U | V | W | X | Y | Z | A | B | C | D | E | F |
| **H** | H | I | J | K | L | M | N | O | P | Q | R | S | T | U | V | W | X | Y | Z | A | B | C | D | E | F | G |
| **I** | I | J | K | L | M | N | O | P | Q | R | S | T | U | V | W | X | Y | Z | A | B | C | D | E | F | G | H |
| **J** | J | K | L | M | N | O | P | Q | R | S | T | U | V | W | X | Y | Z | A | B | C | D | E | F | G | H | I |
| **K** | K | L | M | N | O | P | Q | R | S | T | U | V | W | X | Y | Z | A | B | C | D | E | F | G | H | I | J |
| **L** | L | M | N | O | P | Q | R | S | T | U | V | W | X | Y | Z | A | B | C | D | E | F | G | H | I | J | K |
| **M** | M | N | O | P | Q | R | S | T | U | V | W | X | Y | Z | A | B | C | D | E | F | G | H | I | J | K | L |
| **N** | N | O | P | Q | R | S | T | U | V | W | X | Y | Z | A | B | C | D | E | F | G | H | I | J | K | L | M |
| **O** | O | P | Q | R | S | T | U | V | W | X | Y | Z | A | B | C | D | E | F | G | H | I | J | K | L | M | N |
| **P** | P | Q | R | S | T | U | V | W | X | Y | Z | A | B | C | D | E | F | G | H | I | J | K | L | M | N | O |
| **Q** | Q | R | S | T | U | V | W | X | Y | Z | A | B | C | D | E | F | G | H | I | J | K | L | M | N | O | P |
| **R** | R | S | T | U | V | W | X | Y | Z | A | B | C | D | E | F | G | H | I | J | K | L | M | N | O | P | Q |
| **S** | S | T | U | V | W | X | Y | Z | A | B | C | D | E | F | G | H | I | J | K | L | M | N | O | P | Q | R |
| **T** | T | U | V | W | X | Y | Z | A | B | C | D | E | F | G | H | I | J | K | L | M | N | O | P | Q | R | S |
| **U** | U | V | W | X | Y | Z | A | B | C | D | E | F | G | H | I | J | K | L | M | N | O | P | Q | R | S | T |
| **V** | V | W | X | Y | Z | A | B | C | D | E | F | G | H | I | J | K | L | M | N | O | P | Q | R | S | T | U |
| **W** | W | X | Y | Z | A | B | C | D | E | F | G | H | I | J | K | L | M | N | O | P | Q | R | S | T | U | V |
| **X** | X | Y | Z | A | B | C | D | E | F | G | H | I | J | K | L | M | N | O | P | Q | R | S | T | U | V | W |
| **Y** | Y | Z | A | B | C | D | E | F | G | H | I | J | K | L | M | N | O | P | Q | R | S | T | U | V | W | X |
| **Z** | Z | A | B | C | D | E | F | G | H | I | J | K | L | M | N | O | P | Q | R | S | T | U | V | W | X | Y |

Appendix B

## Ciphering Table: Example

Spirale (Ph.Allard)

| □ | | | | A | B | C | D | E | F | G | H | I | J | K | L | M | N | O | P | Q | R | S | T | U | V | W | X | Y | Z | |
|---|---|---|---|---|---|---|---|---|---|---|---|---|---|---|---|---|---|---|---|---|---|---|---|---|---|---|---|---|---|---|
| new ranks → | | | | 17 | 26 | 21 | 7 | 10 | 3 | 12 | 20 | 14 | 23 | 2 | 15 | 9 | 18 | 6 | 19 | 22 | 25 | 11 | 1 | 16 | 13 | 5 | 4 | 8 | 24 | |
| | ↓ | letters | | 1 | 2 | 3 | 4 | 5 | 6 | 7 | 8 | 9 | 10 | 11 | 12 | 13 | 14 | 15 | 16 | 17 | 18 | 19 | 20 | 21 | 22 | 23 | 24 | 25 | 26 | ↓ rank to letter |
| | | | ↓→ | T | K | F | X | W | O | D | Y | M | E | S | G | V | I | L | U | A | N | P | H | C | Q | J | Z | R | B | |
| A | 13 | 1 | Y | A | B | C | D | E | F | G | H | I | J | K | L | M | N | O | P | Q | R | S | T | U | V | W | X | Y | Z | |
| B | 24 | 2 | Q | B | C | D | E | F | G | H | I | J | K | L | M | N | O | P | Q | R | S | T | U | V | W | X | Y | Z | A | |
| C | 4 | 3 | D | C | D | E | F | G | H | I | J | K | L | M | N | O | P | Q | R | S | T | U | V | W | X | Y | Z | A | B | |
| D | 3 | 4 | C | D | E | F | G | H | I | J | K | L | M | N | O | P | Q | R | S | T | U | V | W | X | Y | Z | A | B | C | |
| E | 17 | 5 | Z | E | F | G | H | I | J | K | L | M | N | O | P | Q | R | S | T | U | V | W | X | Y | Z | A | B | C | D | |
| F | 20 | 6 | W | F | G | H | I | J | K | L | M | N | O | P | Q | R | S | T | U | V | W | X | Y | Z | A | B | C | D | E | |
| G | 23 | 7 | N | G | H | I | J | K | L | M | N | O | P | Q | R | S | T | U | V | W | X | Y | Z | A | B | C | D | E | F | |
| H | 22 | 8 | V | H | I | J | K | L | M | N | O | P | Q | R | S | T | U | V | W | X | Y | Z | A | B | C | D | E | F | G | |
| I | 11 | 9 | U | I | J | K | L | M | N | O | P | Q | R | S | T | U | V | W | X | Y | Z | A | B | C | D | E | F | G | H | |
| J | 26 | 10 | K | J | K | L | M | N | O | P | Q | R | S | T | U | V | W | X | Y | Z | A | B | C | D | E | F | G | H | I | |
| K | 10 | 11 | I | K | L | M | N | O | P | Q | R | S | T | U | V | W | X | Y | Z | A | B | C | D | E | F | G | H | I | J | |
| L | 19 | 12 | T | L | M | N | O | P | Q | R | S | T | U | V | W | X | Y | Z | A | B | C | D | E | F | G | H | I | J | K | |
| M | 18 | 13 | A | M | N | O | P | Q | R | S | T | U | V | W | X | Y | Z | A | B | C | D | E | F | G | H | I | J | K | L | |
| N | 7 | 14 | X | N | O | P | Q | R | S | T | U | V | W | X | Y | Z | A | B | C | D | E | F | G | H | I | J | K | L | M | |
| O | 25 | 15 | S | O | P | Q | R | S | T | U | V | W | X | Y | Z | A | B | C | D | E | F | G | H | I | J | K | L | M | N | |
| P | 16 | 16 | P | P | Q | R | S | T | U | V | W | X | Y | Z | A | B | C | D | E | F | G | H | I | J | K | L | M | N | O | |
| Q | 2 | 17 | E | Q | R | S | T | U | V | W | X | Y | Z | A | B | C | D | E | F | G | H | I | J | K | L | M | N | O | P | |
| R | 21 | 18 | M | R | S | T | U | V | W | X | Y | Z | A | B | C | D | E | F | G | H | I | J | K | L | M | N | O | P | Q | |
| S | 15 | 19 | L | S | T | U | V | W | X | Y | Z | A | B | C | D | E | F | G | H | I | J | K | L | M | N | O | P | Q | R | |
| T | 12 | 20 | F | T | U | V | W | X | Y | Z | A | B | C | D | E | F | G | H | I | J | K | L | M | N | O | P | Q | R | S | |
| U | 9 | 21 | R | U | V | W | X | Y | Z | A | B | C | D | E | F | G | H | I | J | K | L | M | N | O | P | Q | R | S | T | |
| V | 8 | 22 | H | V | W | X | Y | Z | A | B | C | D | E | F | G | H | I | J | K | L | M | N | O | P | Q | R | S | T | U | |
| W | 6 | 23 | G | W | X | Y | Z | A | B | C | D | E | F | G | H | I | J | K | L | M | N | O | P | Q | R | S | T | U | V | |
| X | 14 | 24 | B | X | Y | Z | A | B | C | D | E | F | G | H | I | J | K | L | M | N | O | P | Q | R | S | T | U | V | W | |
| Y | 1 | 25 | O | Y | Z | A | B | C | D | E | F | G | H | I | J | K | L | M | N | O | P | Q | R | S | T | U | V | W | X | |
| Z | 5 | 26 | J | Z | A | B | C | D | E | F | G | H | I | J | K | L | M | N | O | P | Q | R | S | T | U | V | W | X | Y | |

# Permutation Algorithm: Example

Alphabet permutation key

| B | H | M | A | Y | | | |
|---|---|---|---|---|---|---|---|
| 2 | 8 | 13 | 1 | 25 | | | <-- rank in original alphabet |

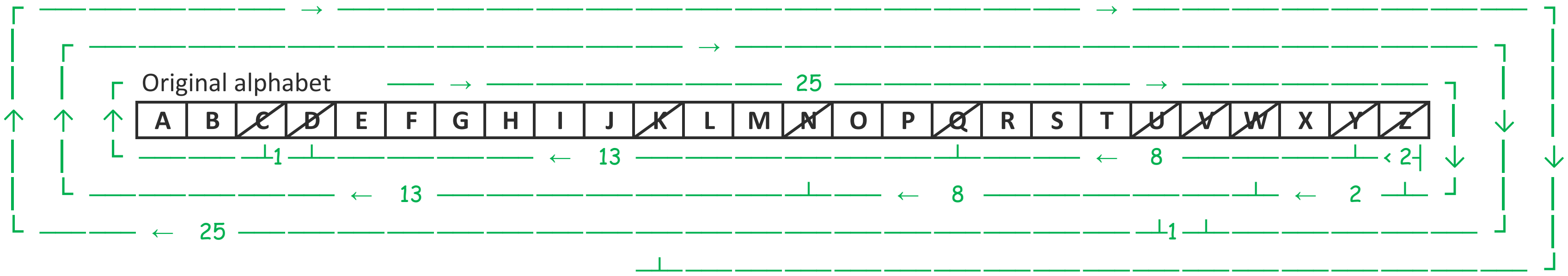

cross out letter in alphabet when it is permuted to jump it later

Permuted alphabet

| | | | | | | | | | | | | | | | | | | | | | | | | | | |
|---|---|---|---|---|---|---|---|---|---|---|---|---|---|---|---|---|---|---|---|---|---|---|---|---|---|---|
| | Y | Q | D | C | Z | W | N | V | U | K | | | | | | | | | | | | | | | | |
| rank in permuted alphabet → | 1 | 2 | 3 | 4 | 5 | 6 | 7 | 8 | 9 | 10 | 11 | 12 | 13 | 14 | 15 | 16 | 17 | 18 | 19 | 20 | 21 | 22 | 23 | 24 | 25 | 26 |
| to search letter → | A | B | C | D | E | F | G | H | I | J | K | L | M | N | O | P | Q | R | S | T | U | V | W | X | Y | Z |
| its rank in permuted alphabet → | | | 4 | 3 | | | | | | | 10 | | | 7 | | | 2 | | | | 9 | 8 | 6 | | 1 | 5 |

Appendix D1: worked example

# Alphabet Permutations

Spirale (Ph.Allard)

Original alphabet

| | | | | | | | | | | | | | | | | | | | | | | | | | |
|---|---|---|---|---|---|---|---|---|---|---|---|---|---|---|---|---|---|---|---|---|---|---|---|---|---|
| A | B | C | ~~D~~ | E | F | ~~G~~ | H | ~~I~~ | J | K | L | ~~M~~ | N | ~~O~~ | P | ~~Q~~ | ~~R~~ | S | T | U | ~~V~~ | ~~W~~ | X | ~~Y~~ | Z |

count in this way for permutation ⟵

after permutation of a letter, cross it in alphabet with a dark pen

Alphabet permutation **Key 1**

| | | | | | | | |
|---|---|---|---|---|---|---|---|
| N | V | I | K | K | I | H | |
| 14 | 22 | 9 | 11 | 11 | 9 | 8 | ⟵ rank in original alphabet |
| .. | .. | .. | . | . | . | . | ⟵ marks to record advancement of permutation process |

Permuted alphabet for **Rows** in Ciphering Table

| | | | | | | | | | | | | | | | | | | | | | | | | | | |
|---|---|---|---|---|---|---|---|---|---|---|---|---|---|---|---|---|---|---|---|---|---|---|---|---|---|---|
| | M | Q | G | V | I | Y | O | W | R | D | L | U | E | P | K | N | T | J | C | A | X | B | S | Z | H | F |
| rank in permuted alphabet → | 1 | 2 | 3 | 4 | 5 | 6 | 7 | 8 | 9 | 10 | 11 | 12 | 13 | 14 | 15 | 16 | 17 | 18 | 19 | 20 | 21 | 22 | 23 | 24 | 25 | 26 |
| to search a letter → | A | B | C | D | E | F | G | H | I | J | K | L | M | N | O | P | Q | R | S | T | U | V | W | X | Y | Z |
| its rank in permuted alphabet → | 20 | 22 | 19 | 10 | 13 | 26 | 3 | 25 | 5 | 18 | 15 | 11 | 1 | 16 | 7 | 14 | 2 | 9 | 23 | 17 | 12 | 4 | 8 | 21 | 6 | 24 |

Original alphabet

| | | | | | | | | | | | | | | | | | | | | | | | | | |
|---|---|---|---|---|---|---|---|---|---|---|---|---|---|---|---|---|---|---|---|---|---|---|---|---|---|
| A | B | C | D | E | F | G | H | I | J | K | L | M | N | O | P | Q | R | S | T | U | V | W | X | Y | Z |

count in this way for permutation ⟵

after permutation of a letter, cross it in alphabet with a dark pen

Alphabet permutation **Key 2**

| | | | | | | | |
|---|---|---|---|---|---|---|---|
| C | T | S | Q | E | O | U | |
| 3 | 20 | 19 | 17 | 5 | 15 | 21 | ⟵ rank in original alphabet |
| | | | | | | | ⟵ marks to record advancement of permutation process |

Permuted alphabet for **Columns** in Ciphering Table

| | | | | | | | | | | | | | | | | | | | | | | | | | | |
|---|---|---|---|---|---|---|---|---|---|---|---|---|---|---|---|---|---|---|---|---|---|---|---|---|---|---|
| | X | D | J | Q | L | T | S | O | M | I | H | B | A | N | F | P | U | W | E | C | V | G | K | Z | Y | R |
| rank in permuted alphabet → | 1 | 2 | 3 | 4 | 5 | 6 | 7 | 8 | 9 | 10 | 11 | 12 | 13 | 14 | 15 | 16 | 17 | 18 | 19 | 20 | 21 | 22 | 23 | 24 | 25 | 26 |
| to search a letter → | A | B | C | D | E | F | G | H | I | J | K | L | M | N | O | P | Q | R | S | T | U | V | W | X | Y | Z |
| its rank in permuted alphabet → | 13 | 12 | 20 | 2 | 19 | 15 | 22 | 11 | 10 | 3 | 23 | 5 | 9 | 14 | 8 | 16 | 4 | 26 | 7 | 6 | 17 | 21 | 18 | 1 | 25 | 24 |

Appendix D2: worked example

## Ciphering Table

Spirale (Ph.Allard)

| | | | | | letter from KEY 4 or KEYSTREAM | | | | | | | | | | | | | | | | | | | | | | | | | | |
|---|---|---|---|---|---|---|---|---|---|---|---|---|---|---|---|---|---|---|---|---|---|---|---|---|---|---|---|---|---|---|---|
| | □ | | | | A | B | C | D | E | F | G | H | I | J | K | L | M | N | O | P | Q | R | S | T | U | V | W | X | Y | Z | |
| | | from new rank to letter | | | 13 | 12 | 20 | 2 | 19 | 15 | 22 | 11 | 10 | 3 | 23 | 5 | 9 | 14 | 8 | 16 | 4 | 26 | 7 | 6 | 17 | 21 | 18 | 1 | 25 | 24 | ┐ from new rank to letter |
| | | | | | 1 | 2 | 3 | 4 | 5 | 6 | 7 | 8 | 9 | 10 | 11 | 12 | 13 | 14 | 15 | 16 | 17 | 18 | 19 | 20 | 21 | 22 | 23 | 24 | 25 | 26 | ↓ |
| | | ┌ | → | ┐ | X | D | J | Q | L | T | S | O | M | I | H | B | A | N | F | P | U | W | E | C | V | G | K | Z | Y | R | ┘ |
| letter from KEY 3 or PLAINTEXT | A | 20 | 1 | M | A | B | C | D | E | F | G | H | I | J | K | L | M | N | O | P | Q | R | S | T | U | V | W | X | Y | Z | |
| | B | 22 | 2 | Q | B | C | D | E | F | G | H | I | J | K | L | M | N | O | P | Q | R | S | T | U | V | W | X | Y | Z | A | |
| | C | 19 | 3 | G | C | D | E | F | G | H | I | J | K | L | M | N | O | P | Q | R | S | T | U | V | W | X | Y | Z | A | B | |
| | D | 10 | 4 | V | D | E | F | G | H | I | J | K | L | M | N | O | P | Q | R | S | T | U | V | W | X | Y | Z | A | B | C | |
| | E | 13 | 5 | I | E | F | G | H | I | J | K | L | M | N | O | P | Q | R | S | T | U | V | W | X | Y | Z | A | B | C | D | |
| | F | 26 | 6 | Y | F | G | H | I | J | K | L | M | N | O | P | Q | R | S | T | U | V | W | X | Y | Z | A | B | C | D | E | |
| | G | 3 | 7 | O | G | H | I | J | K | L | M | N | O | P | Q | R | S | T | U | V | W | X | Y | Z | A | B | C | D | E | F | |
| | H | 25 | 8 | W | H | I | J | K | L | M | N | O | P | Q | R | S | T | U | V | W | X | Y | Z | A | B | C | D | E | F | G | |
| | I | 5 | 9 | R | I | J | K | L | M | N | O | P | Q | R | S | T | U | V | W | X | Y | Z | A | B | C | D | E | F | G | H | |
| | J | 18 | 10 | D | J | K | L | M | N | O | P | Q | R | S | T | U | V | W | X | Y | Z | A | B | C | D | E | F | G | H | I | |
| | K | 15 | 11 | L | K | L | M | N | O | P | Q | R | S | T | U | V | W | X | Y | Z | A | B | C | D | E | F | G | H | I | J | |
| | L | 11 | 12 | U | L | M | N | O | P | Q | R | S | T | U | V | W | X | Y | Z | A | B | C | D | E | F | G | H | I | J | K | |
| | M | 1 | 13 | E | M | N | O | P | Q | R | S | T | U | V | W | X | Y | Z | A | B | C | D | E | F | G | H | I | J | K | L | |
| | N | 16 | 14 | P | N | O | P | Q | R | S | T | U | V | W | X | Y | Z | A | B | C | D | E | F | G | H | I | J | K | L | M | |
| | O | 7 | 15 | K | O | P | Q | R | S | T | U | V | W | X | Y | Z | A | B | C | D | E | F | G | H | I | J | K | L | M | N | |
| | P | 14 | 16 | N | P | Q | R | S | T | U | V | W | X | Y | Z | A | B | C | D | E | F | G | H | I | J | K | L | M | N | O | |
| | Q | 2 | 17 | T | Q | R | S | T | U | V | W | X | Y | Z | A | B | C | D | E | F | G | H | I | J | K | L | M | N | O | P | |
| | R | 9 | 18 | J | R | S | T | U | V | W | X | Y | Z | A | B | C | D | E | F | G | H | I | J | K | L | M | N | O | P | Q | |
| | S | 23 | 19 | C | S | T | U | V | W | X | Y | Z | A | B | C | D | E | F | G | H | I | J | K | L | M | N | O | P | Q | R | |
| | T | 17 | 20 | A | T | U | V | W | X | Y | Z | A | B | C | D | E | F | G | H | I | J | K | L | M | N | O | P | Q | R | S | |
| | U | 12 | 21 | X | U | V | W | X | Y | Z | A | B | C | D | E | F | G | H | I | J | K | L | M | N | O | P | Q | R | S | T | |
| | V | 4 | 22 | B | V | W | X | Y | Z | A | B | C | D | E | F | G | H | I | J | K | L | M | N | O | P | Q | R | S | T | U | |
| | W | 8 | 23 | S | W | X | Y | Z | A | B | C | D | E | F | G | H | I | J | K | L | M | N | O | P | Q | R | S | T | U | V | |
| | X | 21 | 24 | Z | X | Y | Z | A | B | C | D | E | F | G | H | I | J | K | L | M | N | O | P | Q | R | S | T | U | V | W | |
| | Y | 6 | 25 | H | Y | Z | A | B | C | D | E | F | G | H | I | J | K | L | M | N | O | P | Q | R | S | T | U | V | W | X | |
| | Z | 24 | 26 | F | Z | A | B | C | D | E | F | G | H | I | J | K | L | M | N | O | P | Q | R | S | T | U | V | W | X | Y | |

| | | Key 4 | | | | | | |
|---|---|---|---|---|---|---|---|---|
| | | E | A | I | W | D | S | H |
| Key 3 | D | B | V | S | A | K | P | T |
| | N | H | B | Y | G | Q | V | Z |
| | G | U | O | L | T | D | I | M |
| | D | B | V | S | A | K | P | T |
| | K | G | A | X | F | P | U | Y |
| | S | O | I | F | N | X | C | G |
| | Z | P | J | G | O | Y | D | H |

Fill array according with Ciphering Table $Y_{p,q} = X_p \square X_q$

then fill the first 49 cells below by reading array in diagonal from top left corner

and generate the rest of Keystream according $X_p \square X_{p+25} = X_{p+49}$ from p = 1

(copy shaded cells in same rank cells)

| 1 | 2 | 3 | 4 | 5 | 6 | 7 | 8 | 9 | 10 | 11 | 12 | 13 | 14 | 15 | 16 | 17 | 18 | 19 | 20 | 21 | 22 | 23 | 24 | 25 |
|---|---|---|---|---|---|---|---|---|---|---|---|---|---|---|---|---|---|---|---|---|---|---|---|---|
| B | H | V | U | B | S | B | O | Y | A | G | V | L | G | K | O | A | S | T | Q | P | P | I | X | A |

↙

| 26 | 27 | 28 | 29 | 30 | 31 | 32 | 33 | 34 | 35 | 36 | 37 | 38 | 39 | 40 | 41 | 42 | 43 | 44 | 45 | 46 | 47 | 48 | 49 | 50 |
|---|---|---|---|---|---|---|---|---|---|---|---|---|---|---|---|---|---|---|---|---|---|---|---|---|
| D | V | T | J | F | F | K | I | Z | G | N | P | P | M | O | X | U | T | Y | C | Y | D | G | H | W |

| 50 | 51 | 52 | 53 | 54 | 55 | 56 | 57 | 58 | 59 | 60 | 61 | 62 | 63 | 64 | 65 | 66 | 67 | 68 | 69 | 70 | 71 | 72 | 73 | 74 | 75 |
|---|---|---|---|---|---|---|---|---|---|---|---|---|---|---|---|---|---|---|---|---|---|---|---|---|---|
| W | S | I | N | J | K | R | P | C | O | P | S | Z | K | V | G | J | B | O | U | L | O | Z | E | K | P |

| 75 | 76 | 77 | 78 | 79 | 80 | 81 | 82 | 83 | 84 | 85 | 86 | 87 | 88 | 89 | 90 | 91 | 92 | 93 | 94 | 95 | 96 | 97 | 98 | 99 | 100 |
|---|---|---|---|---|---|---|---|---|---|---|---|---|---|---|---|---|---|---|---|---|---|---|---|---|---|
| P | | | | | | | | | | | | | | | | | | | | | | | | | |

| 100 | 101 | 102 | 103 | 104 | 105 | 106 | 107 | 108 | 109 | 110 | 111 | 112 | 113 | 114 | 115 | 116 | 117 | 118 | 119 | 120 | 121 | 122 | 123 | 124 | 125 |
|---|---|---|---|---|---|---|---|---|---|---|---|---|---|---|---|---|---|---|---|---|---|---|---|---|---|
| | | | | | | | | | | | | | | | | | | | | | | | | | |

| 125 | 126 | 127 | 128 | 129 | 130 | 131 | 132 | 133 | 134 | 135 | 136 | 137 | 138 | 139 | 140 | 141 | 142 | 143 | 144 | 145 | 146 | 147 | 148 | 149 | 150 |
|---|---|---|---|---|---|---|---|---|---|---|---|---|---|---|---|---|---|---|---|---|---|---|---|---|---|
| | | | | | | | | | | | | | | | | | | | | | | | | | |

| 150 | 151 | 152 | 153 | 154 | 155 | 156 | 157 | 158 | 159 | 160 | 161 | 162 | 163 | 164 | 165 | 166 | 167 | 168 | 169 | 170 | 171 | 172 | 173 | 174 | 175 |
|---|---|---|---|---|---|---|---|---|---|---|---|---|---|---|---|---|---|---|---|---|---|---|---|---|---|
| | | | | | | | | | | | | | | | | | | | | | | | | | |

| 175 | 176 | 177 | 178 | 179 | 180 | 181 | 182 | 183 | 184 | 185 | 186 | 187 | 188 | 189 | 190 | 191 | 192 | 193 | 194 | 195 | 196 | 197 | 198 | 199 | 200 |
|---|---|---|---|---|---|---|---|---|---|---|---|---|---|---|---|---|---|---|---|---|---|---|---|---|---|
| | | | | | | | | | | | | | | | | | | | | | | | | | |

| 200 | 201 | 202 | 203 | 204 | 205 | 206 | 207 | 208 | 209 | 210 | 211 | 212 | 213 | 214 | 215 | 216 | 217 | 218 | 219 | 220 | 221 | 222 | 223 | 224 | 225 |
|---|---|---|---|---|---|---|---|---|---|---|---|---|---|---|---|---|---|---|---|---|---|---|---|---|---|
| | | | | | | | | | | | | | | | | | | | | | | | | | |

| 225 | 226 | 227 | 228 | 229 | 230 | 231 | 232 | 233 | 234 | 235 | 236 | 237 | 238 | 239 | 240 | 241 | 242 | 243 | 244 | 245 | 246 | 247 | 248 | 249 | 250 |
|---|---|---|---|---|---|---|---|---|---|---|---|---|---|---|---|---|---|---|---|---|---|---|---|---|---|
| | | | | | | | | | | | | | | | | | | | | | | | | | |

| 250 | 251 | 252 | 253 | 254 | 255 | 256 | 257 | 258 | 259 | 260 | 261 | 262 | 263 | 264 | 265 | 266 | 267 | 268 | 269 | 270 | 271 | 272 | 273 | 274 | 275 |
|---|---|---|---|---|---|---|---|---|---|---|---|---|---|---|---|---|---|---|---|---|---|---|---|---|---|
| | | | | | | | | | | | | | | | | | | | | | | | | | |

| 275 | 276 | 277 | 278 | 279 | 280 | 281 | 282 | 283 | 284 | 285 | 286 | 287 | 288 | 289 | 290 | 291 | 292 | 293 | 294 | 295 | 296 | 297 | 298 | 299 | 300 |
|---|---|---|---|---|---|---|---|---|---|---|---|---|---|---|---|---|---|---|---|---|---|---|---|---|---|
| | | | | | | | | | | | | | | | | | | | | | | | | | |

| 300 | 301 | 302 | 303 | 304 | 305 | 306 | 307 | 308 | 309 | 310 | 311 | 312 | 313 | 314 | 315 | 316 | 317 | 318 | 319 | 320 | 321 | 322 | 323 | 324 | 325 |
|---|---|---|---|---|---|---|---|---|---|---|---|---|---|---|---|---|---|---|---|---|---|---|---|---|---|
| | | | | | | | | | | | | | | | | | | | | | | | | | |

| 325 | 326 | 327 | 328 | 329 | 330 | 331 | 332 | 333 | 334 | 335 | 336 | 337 | 338 | 339 | 340 | 341 | 342 | 343 | 344 | 345 | 346 | 347 | 348 | 349 | 350 |
|---|---|---|---|---|---|---|---|---|---|---|---|---|---|---|---|---|---|---|---|---|---|---|---|---|---|
| | | | | | | | | | | | | | | | | | | | | | | | | | |

| 350 | 351 | 352 | 353 | 354 | 355 | 356 | 357 | 358 | 359 | 360 | 361 | 362 | 363 | 364 | 365 | 366 | 367 | 368 | 369 | 370 | 371 | 372 | 373 | 374 | 375 |
|---|---|---|---|---|---|---|---|---|---|---|---|---|---|---|---|---|---|---|---|---|---|---|---|---|---|
| | | | | | | | | | | | | | | | | | | | | | | | | | |

| 375 | 376 | 377 | 378 | 379 | 380 | 381 | 382 | 383 | 384 | 385 | 386 | 387 | 388 | 389 | 390 | 391 | 392 | 393 | 394 | 395 | 396 | 397 | 398 | 399 | 400 |
|---|---|---|---|---|---|---|---|---|---|---|---|---|---|---|---|---|---|---|---|---|---|---|---|---|---|
| | | | | | | | | | | | | | | | | | | | | | | | | | |

Then copy in Appendix D4 the 25-letter segments (without the shaded cells)

## Encryption or Decryption

Encryption ↓ Decryption ↑

| | | | | | | | | | | | | | | | | | | | | | | | | | |
|---|---|---|---|---|---|---|---|---|---|---|---|---|---|---|---|---|---|---|---|---|---|---|---|---|---|
| ↓ Plaintext | S | P | I | R | A | L | E | I | S | A | O | N | E | T | I | M | E | P | A | D | C | R | Y | P | T |
| ↓ Keystream | B | H | V | U | B | S | B | O | Y | A | G | V | L | G | K | O | A | S | T | Q | P | P | I | X | A |
| ↓ Ciphertext | H | X | Y | Y | E | Q | X | L | U | F | B | J | Q | L | A | H | Y | T | Y | M | H | X | O | N | C |

Encryption ↓ Decryption ↑

| | | | | | | | | | | | | | | | | | | | | | | | | | |
|---|---|---|---|---|---|---|---|---|---|---|---|---|---|---|---|---|---|---|---|---|---|---|---|---|---|
| ↓ Plaintext | O | S | Y | S | T | E | M | D | E | S | I | G | N | E | D | T | O | R | E | P | L | A | C | E | S |
| ↓ Keystream | D | V | T | J | F | F | K | I | Z | G | N | P | P | M | O | X | U | T | Y | C | Y | D | G | H | W |
| ↓ Ciphertext | H | Q | K | Y | E | A | W | S | J | R | R | R | E | U | Q | Q | W | N | K | G | I | U | N | W | N |

Encryption ↓ Decryption ↑

| | | | | | | | | | | | | | | | | | | | | | | | | | |
|---|---|---|---|---|---|---|---|---|---|---|---|---|---|---|---|---|---|---|---|---|---|---|---|---|---|
| ↓ Plaintext | O | L | I | T | A | I | R | E | W | H | E | N | O | N | E | H | A | S | N | O | C | A | R | D | S |
| ↓ Keystream | S | I | N | J | K | R | P | C | O | P | S | Z | K | V | G | J | B | O | U | L | O | Z | E | K | P |
| ↓ Ciphertext | M | T | R | S | P | D | X | F | O | N | S | M | C | J | H | A | E | D | F | K | Z | Q | A | F | L |

Encryption ↓ Decryption ↑

| | | | | | | | | | | | | | | | | | | | | | | | | | |
|---|---|---|---|---|---|---|---|---|---|---|---|---|---|---|---|---|---|---|---|---|---|---|---|---|---|
| ↓ Plaintext | | | | | | | | | | | | | | | | | | | | | | | | | |
| ↓ Keystream | | | | | | | | | | | | | | | | | | | | | | | | | |
| ↓ Ciphertext | | | | | | | | | | | | | | | | | | | | | | | | | |

Encryption ↓ Decryption ↑

| | | | | | | | | | | | | | | | | | | | | | | | | | |
|---|---|---|---|---|---|---|---|---|---|---|---|---|---|---|---|---|---|---|---|---|---|---|---|---|---|
| ↓ Plaintext | | | | | | | | | | | | | | | | | | | | | | | | | |
| ↓ Keystream | | | | | | | | | | | | | | | | | | | | | | | | | |
| ↓ Ciphertext | | | | | | | | | | | | | | | | | | | | | | | | | |

Encryption ↓ Decryption ↑

| | | | | | | | | | | | | | | | | | | | | | | | | | |
|---|---|---|---|---|---|---|---|---|---|---|---|---|---|---|---|---|---|---|---|---|---|---|---|---|---|
| ↓ Plaintext | | | | | | | | | | | | | | | | | | | | | | | | | |
| ↓ Keystream | | | | | | | | | | | | | | | | | | | | | | | | | |
| ↓ Ciphertext | | | | | | | | | | | | | | | | | | | | | | | | | |

Encryption ↓ Decryption ↑

| | | | | | | | | | | | | | | | | | | | | | | | | | |
|---|---|---|---|---|---|---|---|---|---|---|---|---|---|---|---|---|---|---|---|---|---|---|---|---|---|
| ↓ Plaintext | | | | | | | | | | | | | | | | | | | | | | | | | |
| ↓ Keystream | | | | | | | | | | | | | | | | | | | | | | | | | |
| ↓ Ciphertext | | | | | | | | | | | | | | | | | | | | | | | | | |

Encryption ↓ Decryption ↑

| | | | | | | | | | | | | | | | | | | | | | | | | | |
|---|---|---|---|---|---|---|---|---|---|---|---|---|---|---|---|---|---|---|---|---|---|---|---|---|---|
| ↓ Plaintext | | | | | | | | | | | | | | | | | | | | | | | | | |
| ↓ Keystream | | | | | | | | | | | | | | | | | | | | | | | | | |
| ↓ Ciphertext | | | | | | | | | | | | | | | | | | | | | | | | | |

# Alphabet Permutations

Original alphabet

| | | | | | | | | | | | | | | | | | | | | | | | | | |
|---|---|---|---|---|---|---|---|---|---|---|---|---|---|---|---|---|---|---|---|---|---|---|---|---|---|
| A | B | C | D | E | F | G | H | I | J | K | L | M | N | O | P | Q | R | S | T | U | V | W | X | Y | Z |

count in this way for permutation ⟵

after permutation of a letter, cross it in alphabet with a dark pen

Alphabet permutation **Key 1**

| | | | | | | | |
|---|---|---|---|---|---|---|---|
| | | | | | | | |
| | | | | | | | ⟵ rank in original alphabet |

⟵ marks to record advancement of permutation process

Permuted alphabet for **Rows** in Ciphering Table

| | | | | | | | | | | | | | | | | | | | | | | | | | | |
|---|---|---|---|---|---|---|---|---|---|---|---|---|---|---|---|---|---|---|---|---|---|---|---|---|---|---|
| | | | | | | | | | | | | | | | | | | | | | | | | | | |
| rank in permuted alphabet ⟶ | **1** | **2** | **3** | **4** | **5** | **6** | **7** | **8** | **9** | **10** | **11** | **12** | **13** | **14** | **15** | **16** | **17** | **18** | **19** | **20** | **21** | **22** | **23** | **24** | **25** | **26** |
| to search a letter ⟶ | A | B | C | D | E | F | G | H | I | J | K | L | M | N | O | P | Q | R | S | T | U | V | W | X | Y | Z |
| its rank in permuted alphabet ⟶ | | | | | | | | | | | | | | | | | | | | | | | | | | |

Original alphabet

| | | | | | | | | | | | | | | | | | | | | | | | | | |
|---|---|---|---|---|---|---|---|---|---|---|---|---|---|---|---|---|---|---|---|---|---|---|---|---|---|
| A | B | C | D | E | F | G | H | I | J | K | L | M | N | O | P | Q | R | S | T | U | V | W | X | Y | Z |

count in this way for permutation ⟵

after permutation of a letter, cross it in alphabet with a dark pen

Alphabet permutation **Key 2**

| | | | | | | | |
|---|---|---|---|---|---|---|---|
| | | | | | | | |
| | | | | | | | ⟵ rank in original alphabet |

⟵ marks to record advancement of permutation process

Permuted alphabet for **Columns** in Ciphering Table

| | | | | | | | | | | | | | | | | | | | | | | | | | | |
|---|---|---|---|---|---|---|---|---|---|---|---|---|---|---|---|---|---|---|---|---|---|---|---|---|---|---|
| | | | | | | | | | | | | | | | | | | | | | | | | | | |
| rank in permuted alphabet ⟶ | **1** | **2** | **3** | **4** | **5** | **6** | **7** | **8** | **9** | **10** | **11** | **12** | **13** | **14** | **15** | **16** | **17** | **18** | **19** | **20** | **21** | **22** | **23** | **24** | **25** | **26** |
| to search a letter ⟶ | A | B | C | D | E | F | G | H | I | J | K | L | M | N | O | P | Q | R | S | T | U | V | W | X | Y | Z |
| its rank in permuted alphabet ⟶ | | | | | | | | | | | | | | | | | | | | | | | | | | |

## Ciphering Table

letter from KEY 4 or KEYSTREAM

letter from KEY 3 or PLAINTEXT

| □ | | | | A | B | C | D | E | F | G | H | I | J | K | L | M | N | O | P | Q | R | S | T | U | V | W | X | Y | Z |
|---|---|---|---|---|---|---|---|---|---|---|---|---|---|---|---|---|---|---|---|---|---|---|---|---|---|---|---|---|---|
| new ranks → | | | | | | | | | | | | | | | | | | | | | | | | | | | | | |
| ↓ letters | | | | 1 | 2 | 3 | 4 | 5 | 6 | 7 | 8 | 9 | 10 | 11 | 12 | 13 | 14 | 15 | 16 | 17 | 18 | 19 | 20 | 21 | 22 | 23 | 24 | 25 | 26 |
| ┌ → ↓→ | | | | | | | | | | | | | | | | | | | | | | | | | | | | | |
| A | | 1 | | A | B | C | D | E | F | G | H | I | J | K | L | M | N | O | P | Q | R | S | T | U | V | W | X | Y | Z |
| B | | 2 | | B | C | D | E | F | G | H | I | J | K | L | M | N | O | P | Q | R | S | T | U | V | W | X | Y | Z | A |
| C | | 3 | | C | D | E | F | G | H | I | J | K | L | M | N | O | P | Q | R | S | T | U | V | W | X | Y | Z | A | B |
| D | | 4 | | D | E | F | G | H | I | J | K | L | M | N | O | P | Q | R | S | T | U | V | W | X | Y | Z | A | B | C |
| E | | 5 | | E | F | G | H | I | J | K | L | M | N | O | P | Q | R | S | T | U | V | W | X | Y | Z | A | B | C | D |
| F | | 6 | | F | G | H | I | J | K | L | M | N | O | P | Q | R | S | T | U | V | W | X | Y | Z | A | B | C | D | E |
| G | | 7 | | G | H | I | J | K | L | M | N | O | P | Q | R | S | T | U | V | W | X | Y | Z | A | B | C | D | E | F |
| H | | 8 | | H | I | J | K | L | M | N | O | P | Q | R | S | T | U | V | W | X | Y | Z | A | B | C | D | E | F | G |
| I | | 9 | | I | J | K | L | M | N | O | P | Q | R | S | T | U | V | W | X | Y | Z | A | B | C | D | E | F | G | H |
| J | | 10 | | J | K | L | M | N | O | P | Q | R | S | T | U | V | W | X | Y | Z | A | B | C | D | E | F | G | H | I |
| K | | 11 | | K | L | M | N | O | P | Q | R | S | T | U | V | W | X | Y | Z | A | B | C | D | E | F | G | H | I | J |
| L | | 12 | | L | M | N | O | P | Q | R | S | T | U | V | W | X | Y | Z | A | B | C | D | E | F | G | H | I | J | K |
| M | | 13 | | M | N | O | P | Q | R | S | T | U | V | W | X | Y | Z | A | B | C | D | E | F | G | H | I | J | K | L |
| N | | 14 | | N | O | P | Q | R | S | T | U | V | W | X | Y | Z | A | B | C | D | E | F | G | H | I | J | K | L | M |
| O | | 15 | | O | P | Q | R | S | T | U | V | W | X | Y | Z | A | B | C | D | E | F | G | H | I | J | K | L | M | N |
| P | | 16 | | P | Q | R | S | T | U | V | W | X | Y | Z | A | B | C | D | E | F | G | H | I | J | K | L | M | N | O |
| Q | | 17 | | Q | R | S | T | U | V | W | X | Y | Z | A | B | C | D | E | F | G | H | I | J | K | L | M | N | O | P |
| R | | 18 | | R | S | T | U | V | W | X | Y | Z | A | B | C | D | E | F | G | H | I | J | K | L | M | N | O | P | Q |
| S | | 19 | | S | T | U | V | W | X | Y | Z | A | B | C | D | E | F | G | H | I | J | K | L | M | N | O | P | Q | R |
| T | | 20 | | T | U | V | W | X | Y | Z | A | B | C | D | E | F | G | H | I | J | K | L | M | N | O | P | Q | R | S |
| U | | 21 | | U | V | W | X | Y | Z | A | B | C | D | E | F | G | H | I | J | K | L | M | N | O | P | Q | R | S | T |
| V | | 22 | | V | W | X | Y | Z | A | B | C | D | E | F | G | H | I | J | K | L | M | N | O | P | Q | R | S | T | U |
| W | | 23 | | W | X | Y | Z | A | B | C | D | E | F | G | H | I | J | K | L | M | N | O | P | Q | R | S | T | U | V |
| X | | 24 | | X | Y | Z | A | B | C | D | E | F | G | H | I | J | K | L | M | N | O | P | Q | R | S | T | U | V | W |
| Y | | 25 | | Y | Z | A | B | C | D | E | F | G | H | I | J | K | L | M | N | O | P | Q | R | S | T | U | V | W | X |
| Z | | 26 | | Z | A | B | C | D | E | F | G | H | I | J | K | L | M | N | O | P | Q | R | S | T | U | V | W | X | Y |

┐ from new rank to ↓ letter ┘

## Long Key Generating --- Keystream Generating

| | Key 4 | | | | | | |
|---|---|---|---|---|---|---|---|
| | | | | | | | |
| Key 3 | | | | | | | |
| | | | | | | | |
| | | | | | | | |
| | | | | | | | |
| | | | | | | | |
| | | | | | | | |
| | | | | | | | |

Fill array according $Y_{p,q} = X_p \square X_q$
with Ciphering Table

then fill the first 49 cells below by reading array in diagonal from top left corner

and generate the rest of Keystream according $X_p \square X_{p+25} = X_{p+49}$ from p = 1

(copy shaded cells in same rank cells)

| | | | | | | | | | | | | | | | | | | | | | | | | | |
|---|---|---|---|---|---|---|---|---|---|---|---|---|---|---|---|---|---|---|---|---|---|---|---|---|---|
| | 1 | 2 | 3 | 4 | 5 | 6 | 7 | 8 | 9 | 10 | 11 | 12 | 13 | 14 | 15 | 16 | 17 | 18 | 19 | 20 | 21 | 22 | 23 | 24 | 25 |
| | | | | | | | | | | | | | | | | | | | | | | | | | |
| | 26 | 27 | 28 | 29 | 30 | 31 | 32 | 33 | 34 | 35 | 36 | 37 | 38 | 39 | 40 | 41 | 42 | 43 | 44 | 45 | 46 | 47 | 48 | 49 | 50 |
| ↙ | | | | | | | | | | | | | | | | | | | | | | | | | |
| 50 | 51 | 52 | 53 | 54 | 55 | 56 | 57 | 58 | 59 | 60 | 61 | 62 | 63 | 64 | 65 | 66 | 67 | 68 | 69 | 70 | 71 | 72 | 73 | 74 | 75 |
| | | | | | | | | | | | | | | | | | | | | | | | | | |
| 75 | 76 | 77 | 78 | 79 | 80 | 81 | 82 | 83 | 84 | 85 | 86 | 87 | 88 | 89 | 90 | 91 | 92 | 93 | 94 | 95 | 96 | 97 | 98 | 99 | 100 |
| | | | | | | | | | | | | | | | | | | | | | | | | | |
| 100 | 101 | 102 | 103 | 104 | 105 | 106 | 107 | 108 | 109 | 110 | 111 | 112 | 113 | 114 | 115 | 116 | 117 | 118 | 119 | 120 | 121 | 122 | 123 | 124 | 125 |
| | | | | | | | | | | | | | | | | | | | | | | | | | |
| 125 | 126 | 127 | 128 | 129 | 130 | 131 | 132 | 133 | 134 | 135 | 136 | 137 | 138 | 139 | 140 | 141 | 142 | 143 | 144 | 145 | 146 | 147 | 148 | 149 | 150 |
| | | | | | | | | | | | | | | | | | | | | | | | | | |
| 150 | 151 | 152 | 153 | 154 | 155 | 156 | 157 | 158 | 159 | 160 | 161 | 162 | 163 | 164 | 165 | 166 | 167 | 168 | 169 | 170 | 171 | 172 | 173 | 174 | 175 |
| | | | | | | | | | | | | | | | | | | | | | | | | | |
| 175 | 176 | 177 | 178 | 179 | 180 | 181 | 182 | 183 | 184 | 185 | 186 | 187 | 188 | 189 | 190 | 191 | 192 | 193 | 194 | 195 | 196 | 197 | 198 | 199 | 200 |
| | | | | | | | | | | | | | | | | | | | | | | | | | |
| 200 | 201 | 202 | 203 | 204 | 205 | 206 | 207 | 208 | 209 | 210 | 211 | 212 | 213 | 214 | 215 | 216 | 217 | 218 | 219 | 220 | 221 | 222 | 223 | 224 | 225 |
| | | | | | | | | | | | | | | | | | | | | | | | | | |
| 225 | 226 | 227 | 228 | 229 | 230 | 231 | 232 | 233 | 234 | 235 | 236 | 237 | 238 | 239 | 240 | 241 | 242 | 243 | 244 | 245 | 246 | 247 | 248 | 249 | 250 |
| | | | | | | | | | | | | | | | | | | | | | | | | | |
| 250 | 251 | 252 | 253 | 254 | 255 | 256 | 257 | 258 | 259 | 260 | 261 | 262 | 263 | 264 | 265 | 266 | 267 | 268 | 269 | 270 | 271 | 272 | 273 | 274 | 275 |
| | | | | | | | | | | | | | | | | | | | | | | | | | |
| 275 | 276 | 277 | 278 | 279 | 280 | 281 | 282 | 283 | 284 | 285 | 286 | 287 | 288 | 289 | 290 | 291 | 292 | 293 | 294 | 295 | 296 | 297 | 298 | 299 | 300 |
| | | | | | | | | | | | | | | | | | | | | | | | | | |
| 300 | 301 | 302 | 303 | 304 | 305 | 306 | 307 | 308 | 309 | 310 | 311 | 312 | 313 | 314 | 315 | 316 | 317 | 318 | 319 | 320 | 321 | 322 | 323 | 324 | 325 |
| | | | | | | | | | | | | | | | | | | | | | | | | | |
| 325 | 326 | 327 | 328 | 329 | 330 | 331 | 332 | 333 | 334 | 335 | 336 | 337 | 338 | 339 | 340 | 341 | 342 | 343 | 344 | 345 | 346 | 347 | 348 | 349 | 350 |
| | | | | | | | | | | | | | | | | | | | | | | | | | |
| 350 | 351 | 352 | 353 | 354 | 355 | 356 | 357 | 358 | 359 | 360 | 361 | 362 | 363 | 364 | 365 | 366 | 367 | 368 | 369 | 370 | 371 | 372 | 373 | 374 | 375 |
| | | | | | | | | | | | | | | | | | | | | | | | | | |
| 375 | 376 | 377 | 378 | 379 | 380 | 381 | 382 | 383 | 384 | 385 | 386 | 387 | 388 | 389 | 390 | 391 | 392 | 393 | 394 | 395 | 396 | 397 | 398 | 399 | 400 |
| | | | | | | | | | | | | | | | | | | | | | | | | | |

Then copy in Appendix E4 the 25-letter segments (without the shaded cells)

Appendix E3a
for right-to-left writings

## **Long Key Generating --- Keystream Generating**

Spirale (Ph.Allard)

| | | Key 4 | | | | | | |
|---|---|---|---|---|---|---|---|---|
| | | | | | | | | |
| Key 3 | | | | | | | | |
| | | | | | | | | |
| | | | | | | | | |
| | | | | | | | | |
| | | | | | | | | |
| | | | | | | | | |
| | | | | | | | | |

Fill array according with Ciphering Table $Y_{p,q} = X_p \square X_q$

then fill the first 49 cells below by reading array in diagonal from top left corner

and generate the rest of Keystream according $X_p \square X_{p+25} = X_{p+49}$ from p = 1

(copy shaded cells in same rank cells)

| 25 | 24 | 23 | 22 | 21 | 20 | 19 | 18 | 17 | 16 | 15 | 14 | 13 | 12 | 11 | 10 | 9 | 8 | 7 | 6 | 5 | 4 | 3 | 2 | 1 |
|---|---|---|---|---|---|---|---|---|---|---|---|---|---|---|---|---|---|---|---|---|---|---|---|---|
| | | | | | | | | | | | | | | | | | | | | | | | | |

| 50 | 49 | 48 | 47 | 46 | 45 | 44 | 43 | 42 | 41 | 40 | 39 | 38 | 37 | 36 | 35 | 34 | 33 | 32 | 31 | 30 | 29 | 28 | 27 | 26 | |
|---|---|---|---|---|---|---|---|---|---|---|---|---|---|---|---|---|---|---|---|---|---|---|---|---|---|
| | | | | | | | | | | | | | | | | | | | | | | | | | ↘ |

| 75 | 74 | 73 | 72 | 71 | 70 | 69 | 68 | 67 | 66 | 65 | 64 | 63 | 62 | 61 | 60 | 59 | 58 | 57 | 56 | 55 | 54 | 53 | 52 | 51 | 50 |
|---|---|---|---|---|---|---|---|---|---|---|---|---|---|---|---|---|---|---|---|---|---|---|---|---|---|
| | | | | | | | | | | | | | | | | | | | | | | | | | |

| 100 | 99 | 98 | 97 | 96 | 95 | 94 | 93 | 92 | 91 | 90 | 89 | 88 | 87 | 86 | 85 | 84 | 83 | 82 | 81 | 80 | 79 | 78 | 77 | 76 | 75 |
|---|---|---|---|---|---|---|---|---|---|---|---|---|---|---|---|---|---|---|---|---|---|---|---|---|---|
| | | | | | | | | | | | | | | | | | | | | | | | | | |

| 125 | 124 | 123 | 122 | 121 | 120 | 119 | 118 | 117 | 116 | 115 | 114 | 113 | 112 | 111 | 110 | 109 | 108 | 107 | 106 | 105 | 104 | 103 | 102 | 101 | 100 |
|---|---|---|---|---|---|---|---|---|---|---|---|---|---|---|---|---|---|---|---|---|---|---|---|---|---|
| | | | | | | | | | | | | | | | | | | | | | | | | | |

| 150 | 149 | 148 | 147 | 146 | 145 | 144 | 143 | 142 | 141 | 140 | 139 | 138 | 137 | 136 | 135 | 134 | 133 | 132 | 131 | 130 | 129 | 128 | 127 | 126 | 125 |
|---|---|---|---|---|---|---|---|---|---|---|---|---|---|---|---|---|---|---|---|---|---|---|---|---|---|
| | | | | | | | | | | | | | | | | | | | | | | | | | |

| 175 | 174 | 173 | 172 | 171 | 170 | 169 | 168 | 167 | 166 | 165 | 164 | 163 | 162 | 161 | 160 | 159 | 158 | 157 | 156 | 155 | 154 | 153 | 152 | 151 | 150 |
|---|---|---|---|---|---|---|---|---|---|---|---|---|---|---|---|---|---|---|---|---|---|---|---|---|---|
| | | | | | | | | | | | | | | | | | | | | | | | | | |

| 200 | 199 | 198 | 197 | 196 | 195 | 194 | 193 | 192 | 191 | 190 | 189 | 188 | 187 | 186 | 185 | 184 | 183 | 182 | 181 | 180 | 179 | 178 | 177 | 176 | 175 |
|---|---|---|---|---|---|---|---|---|---|---|---|---|---|---|---|---|---|---|---|---|---|---|---|---|---|
| | | | | | | | | | | | | | | | | | | | | | | | | | |

| 225 | 224 | 223 | 222 | 221 | 220 | 219 | 218 | 217 | 216 | 215 | 214 | 213 | 212 | 211 | 210 | 209 | 208 | 207 | 206 | 205 | 204 | 203 | 202 | 201 | 200 |
|---|---|---|---|---|---|---|---|---|---|---|---|---|---|---|---|---|---|---|---|---|---|---|---|---|---|
| | | | | | | | | | | | | | | | | | | | | | | | | | |

| 250 | 249 | 248 | 247 | 246 | 245 | 244 | 243 | 242 | 241 | 240 | 239 | 238 | 237 | 236 | 235 | 234 | 233 | 232 | 231 | 230 | 229 | 228 | 227 | 226 | 225 |
|---|---|---|---|---|---|---|---|---|---|---|---|---|---|---|---|---|---|---|---|---|---|---|---|---|---|
| | | | | | | | | | | | | | | | | | | | | | | | | | |

| 275 | 274 | 273 | 272 | 271 | 270 | 269 | 268 | 267 | 266 | 265 | 264 | 263 | 262 | 261 | 260 | 259 | 258 | 257 | 256 | 255 | 254 | 253 | 252 | 251 | 250 |
|---|---|---|---|---|---|---|---|---|---|---|---|---|---|---|---|---|---|---|---|---|---|---|---|---|---|
| | | | | | | | | | | | | | | | | | | | | | | | | | |

| 300 | 299 | 298 | 297 | 296 | 295 | 294 | 293 | 292 | 291 | 290 | 289 | 288 | 287 | 286 | 285 | 284 | 283 | 282 | 281 | 280 | 279 | 278 | 277 | 276 | 275 |
|---|---|---|---|---|---|---|---|---|---|---|---|---|---|---|---|---|---|---|---|---|---|---|---|---|---|
| | | | | | | | | | | | | | | | | | | | | | | | | | |

| 325 | 324 | 323 | 322 | 321 | 320 | 319 | 318 | 317 | 316 | 315 | 314 | 313 | 312 | 311 | 310 | 309 | 308 | 307 | 306 | 305 | 304 | 303 | 302 | 301 | 300 |
|---|---|---|---|---|---|---|---|---|---|---|---|---|---|---|---|---|---|---|---|---|---|---|---|---|---|
| | | | | | | | | | | | | | | | | | | | | | | | | | |

| 350 | 349 | 348 | 347 | 346 | 345 | 344 | 343 | 342 | 341 | 340 | 339 | 338 | 337 | 336 | 335 | 334 | 333 | 332 | 331 | 330 | 329 | 328 | 327 | 326 | 325 |
|---|---|---|---|---|---|---|---|---|---|---|---|---|---|---|---|---|---|---|---|---|---|---|---|---|---|
| | | | | | | | | | | | | | | | | | | | | | | | | | |

| 375 | 374 | 373 | 372 | 371 | 370 | 369 | 368 | 367 | 366 | 365 | 364 | 363 | 362 | 361 | 360 | 359 | 358 | 357 | 356 | 355 | 354 | 353 | 352 | 351 | 350 |
|---|---|---|---|---|---|---|---|---|---|---|---|---|---|---|---|---|---|---|---|---|---|---|---|---|---|
| | | | | | | | | | | | | | | | | | | | | | | | | | |

| 400 | 399 | 398 | 397 | 396 | 395 | 394 | 393 | 392 | 391 | 390 | 389 | 388 | 387 | 386 | 385 | 384 | 383 | 382 | 381 | 380 | 379 | 378 | 377 | 376 | 375 |
|---|---|---|---|---|---|---|---|---|---|---|---|---|---|---|---|---|---|---|---|---|---|---|---|---|---|
| | | | | | | | | | | | | | | | | | | | | | | | | | |

Then copy in Appendix E4 the 25-letter segments (without the shaded cells)

# Encryption or Decryption

↓ Plaintext
Encryption ↓
↓ Keystream
↓ Ciphertext
Decryption ↑

↓ Plaintext
Encryption ↓
↓ Keystream
↓ Ciphertext
Decryption ↑

↓ Plaintext
Encryption ↓
↓ Keystream
↓ Ciphertext
Decryption ↑

↓ Plaintext
Encryption ↓
↓ Keystream
↓ Ciphertext
Decryption ↑

↓ Plaintext
Encryption ↓
↓ Keystream
↓ Ciphertext
Decryption ↑

↓ Plaintext
Encryption ↓
↓ Keystream
↓ Ciphertext
Decryption ↑

↓ Plaintext
Encryption ↓
↓ Keystream
↓ Ciphertext
Decryption ↑

↓ Plaintext
Encryption ↓
↓ Keystream
↓ Ciphertext
Decryption ↑

**Ciphertext 1**

XEXEQPVDKYMVCCRZTMRLCCKQBKPOEVZXYQDDCIOEINTLVQJKATRBDWEEMVMYOOEIOOVMOCRSBJGSNZUQJZT
XODHAOTRIEJRPENVKDJYVLNPOERZSFZTIHTZJMMOTBGRJCCZMVOUNWMKTPCPPCASJFAVUEJPTJRTWFCHIGZTGT
CJEYRZDISQEKTKPNIBNAPQSUKCUPWKSZBSNOATKXKGHRAMONICEEGJBZGBLRFRHBYTHITNLXRFCZPLZOEUTBMJO
EGLLEHSNBAYNWONHAQVFDEHLTANKZIQTVZXQFOXHQJTMWHMDWFKMPEQJ

**Ciphertext 2**

BDXVLCJKYIFKACDRUUDLVCOEVKYUJPISHQJYFFKNDGEPWIVJRQKCLJJSZPZMEKHLJHVXXEQUNGTSNJPMRCDXNIQG
KCBBTULFBCFYUSYTZSOHFMLXXEDZINIOQBYQIGDULEBRNFQUUGOSYGNFQOSRHGZWZIHINHTKTIKGTAPIYDTXHA
POXJDKEEMENNFORRCBFPXZPELBEGTNCWWNYIFMCFNECDTZHHEHHDITXKFYXROXETFWZMYCIVNPLEEAAIEYSSM
BIFWSPLVUXZBCHEIDZEXJLRQYDULAPZSBFBGSWYTCTVPTYDWIJHYSJNSNGYIWNUBWGRMCHAEWXOAFLQOUUKS
OOEFDFKRFKJSUUVUSZNVYARBGSBTWHVAUVSQVJLNTWHBIMGTWVRCWOPIZHGWJSZNKIJYTERPKWAJTCWTJEU
PVRJGRBFEGDINAITTHIWAOZDUXOKOWSHKYPEYJANJUUMNRZQOWTOUDMZQIXIGTCVHSAHJGFDSFXGSIMTLEB
WGIMTEOZINRXFLWKKSVQFFFDMEMNEVCIZPYZGYLAXIRHRDFASPDIVRTVQFJLPBVEPJEPYJRIIZXWASMYEIRETRAFS
EEOYQARNQMLONRFVMAXQSQGUAVVZYUYUNAGGTOSJGIRFISJS

**Ciphertext 3**

IALVBQTIJDXJUDGFJTOCLLFDXJLVFCKNZDBKNGLNYIPPXBEOMKHQKDWKBGKFVBOINSHJTZBSUUGDMPUOVRQVV
GBONKTSLKZABMHKVKLOKVTGDJGCVDYVBKOOMHQQAUAGDSTGMLQGEOFTVSRUBJPAJMHBUPXNQDUHCCWDF
UOUJRJXQWZFQONUCNKKKRYAOMINZFKBDNAJKCOFVQHJRONKJJSMVEFKHNLTGUYEPAOIAZSWCIYGJPAAQZYUN
QVJECPNGVWFUGKMUMHFTQGRBNVTSVKOSTKQFSTSCNEIQJMGANCXNWFKCPSWYFYNLBYFDYVBQXKLTFSOPHF
NSDBVMCDUJMIGMUJELPCCGFCAQNCKYHUARBJDXPGMHKLFYZLJFCFXQQHQFMCANVWHOHJPHPIHSVIJQXNCMF
IUMWAFRSEISJMIJYNAIOIKMBURSAMQOMNZXHYJKHZGPYXNRIIYTHVCSMEKEBVQSFIDNSEPABWDRXMWEXFQW
OLKOSTABYGUETEYQIJIKXPTEBDBFBCATVAPQNYOFORTUATIOGCRYRCACNJJXQVQEBNQCGSDMAEELVZBFAZCNNN
LHHYNWHVHDHAZRKHZJWBTRDQIKBGXYGEHSVBHQPDHLMYKKGIEKELKAXAYICZZFHLJAWMVRCXPHVYKMNMUD
UOEDEBNJEPTSVSRQWDZNGAJJOTVMZWHWOOLDKJAFONANCZMODJWUKKXWDYNJXZKBYOOOWVOZGYAGVOT
EWUYRAAPDPNEOLBXFYFEMNCUCINEHXMDDSAYTUEXQCCOWXXVLFPAIZTIWYLDHAOYIZDDYGKRJWJOUSMCBNU
AOOUKJKCPPJUJDTYIFTJJPLLXDMMZUDSBJYQRPUCZJWAKOEGKOOMNCBBNSBGCSMYCJQHOCBNKUOLHBCCLRXJJ
AOMPJBIRBMVEGHODVICBTZFYWAIYQTKCICSWOIHVPPANRWRFFJAS

**Ciphertext 4**

FULDQYOBJCEQPSDHNLYLYBOXOIIFJCLHJLCIBBZKJBDIPLGMDTRIQRLZIXJVCQJXFJZVVHBVLNNKDWKFRKVDMVQEI
GMTAQVZQYCGVEMNWTQGBBCASKZHMFDUUYVPAXNJMVVVFCGRRPDNHLVMQJLKLRAIWWYBQWYBXNNQYDAX
QPCHEDDTEKBOEPUDOBMYDYGNPXJIXEBYLTUIMGNTJNISDAGOQEPBZUNAPFPRXFSUWTOQYZLZUHTHZNOCRW
WPCJQBMACNXFINPNDHOTQQFCEQDZBRREZSJIXNVGFJMHGKEVGVKTQQAZGOENOTJZUPWGHKDSGDHAHFHHTU
RJDPDDHIHWEMOBGHWHQGCJXSOGGUPQNCTWUDIFJRXCNHABXOTJFORNMIFBPZRNRPSLBTDNMFONQDFQSYN
LWWLEUTCUTSHTRKAKMVNWTGQALUADQHZGDHKCNTFVBXWLMQIQMCOPVHIXOGUJIFIOFTRKRWO

# Extended Alphabet Permutations

Spirale (Ph.Allard)

Original alphabet

| A | B | C | D | E | F | G | H | I | J | K | L | M | N | O | P | Q | R | S | T | U | V | W | X | Y | Z | 0 | 1 | 2 | 3 | 4 | 5 | 6 | 7 | 8 | 9 |
|---|---|---|---|---|---|---|---|---|---|---|---|---|---|---|---|---|---|---|---|---|---|---|---|---|---|---|---|---|---|---|---|---|---|---|---|

Alphabet permutation **Key 1**

| | | | | | | |
|---|---|---|---|---|---|---|
| | | | | | | |
| | | | | | | |

← rank in original alphabet
← marks to record advancement of permutation process

Permuted alphabet for **Rows** in Ciphering Table

| | | | | | | | | | | | | | | | | | | | | | | | | | | | | | | | | | | | | |
|---|---|---|---|---|---|---|---|---|---|---|---|---|---|---|---|---|---|---|---|---|---|---|---|---|---|---|---|---|---|---|---|---|---|---|---|---|
| character | | | | | | | | | | | | | | | | | | | | | | | | | | | | | | | | | | | | |
| | **1** | **2** | **3** | **4** | **5** | **6** | **7** | **8** | **9** | **10** | **11** | **12** | **13** | **14** | **15** | **16** | **17** | **18** | **19** | **20** | **21** | **22** | **23** | **24** | **25** | **26** | **27** | **28** | **29** | **30** | **31** | **32** | **33** | **34** | **35** | **36** |
| | **A** | **B** | **C** | **D** | **E** | **F** | **G** | **H** | **I** | **J** | **K** | **L** | **M** | **N** | **O** | **P** | **Q** | **R** | **S** | **T** | **U** | **V** | **W** | **X** | **Y** | **Z** | **0** | **1** | **2** | **3** | **4** | **5** | **6** | **7** | **8** | **9** |
| rank | | | | | | | | | | | | | | | | | | | | | | | | | | | | | | | | | | | | |

Original alphabet

| A | B | C | D | E | F | G | H | I | J | K | L | M | N | O | P | Q | R | S | T | U | V | W | X | Y | Z | 0 | 1 | 2 | 3 | 4 | 5 | 6 | 7 | 8 | 9 |
|---|---|---|---|---|---|---|---|---|---|---|---|---|---|---|---|---|---|---|---|---|---|---|---|---|---|---|---|---|---|---|---|---|---|---|---|

Alphabet permutation **Key 2**

| | | | | | | |
|---|---|---|---|---|---|---|
| | | | | | | |
| | | | | | | |

← rank in original alphabet
← marks to record advancement of permutation process

Permuted alphabet for **Columns** in Ciphering Table

| | | | | | | | | | | | | | | | | | | | | | | | | | | | | | | | | | | | | |
|---|---|---|---|---|---|---|---|---|---|---|---|---|---|---|---|---|---|---|---|---|---|---|---|---|---|---|---|---|---|---|---|---|---|---|---|---|
| character | | | | | | | | | | | | | | | | | | | | | | | | | | | | | | | | | | | | |
| | **1** | **2** | **3** | **4** | **5** | **6** | **7** | **8** | **9** | **10** | **11** | **12** | **13** | **14** | **15** | **16** | **17** | **18** | **19** | **20** | **21** | **22** | **23** | **24** | **25** | **26** | **27** | **28** | **29** | **30** | **31** | **32** | **33** | **34** | **35** | **36** |
| | **A** | **B** | **C** | **D** | **E** | **F** | **G** | **H** | **I** | **J** | **K** | **L** | **M** | **N** | **O** | **P** | **Q** | **R** | **S** | **T** | **U** | **V** | **W** | **X** | **Y** | **Z** | **0** | **1** | **2** | **3** | **4** | **5** | **6** | **7** | **8** | **9** |
| rank | | | | | | | | | | | | | | | | | | | | | | | | | | | | | | | | | | | | |

# Extended Alphabet Permutations

| | | | | | | | | | | | | | | | | |
|---|---|---|---|---|---|---|---|---|---|---|---|---|---|---|---|---|
| | , | . | ( | ) | + | - | * | / | ^ | < | = | > | % | € | ₤ | $ |

count in this way for permutation ⟵

| | | | | | | | | | | | | | | | | |
|---|---|---|---|---|---|---|---|---|---|---|---|---|---|---|---|---|
| | | | | | | | | | | | | | | | | |
| 37 | 38 | 39 | 40 | 41 | 42 | 43 | 44 | 45 | 46 | 47 | 48 | 49 | 50 | 51 | 52 | 53 |
| | , | . | ( | ) | + | - | * | / | ^ | < | = | > | % | € | ₤ | $ |
| | | | | | | | | | | | | | | | | |
| | , | . | ( | ) | + | - | * | / | ^ | < | = | > | % | € | ₤ | $ |

count in this way for permutation ⟵

| | | | | | | | | | | | | | | | | |
|---|---|---|---|---|---|---|---|---|---|---|---|---|---|---|---|---|
| | | | | | | | | | | | | | | | | |
| 37 | 38 | 39 | 40 | 41 | 42 | 43 | 44 | 45 | 46 | 47 | 48 | 49 | 50 | 51 | 52 | 53 |
| | , | . | ( | ) | + | - | * | / | ^ | < | = | > | % | € | ₤ | $ |
| | | | | | | | | | | | | | | | | |

Appendix G2

## Ciphering Table

Spirale (Ph.Allard)

character from KEY 4 or KEYSTREAM

character from KEY 3 or PLAINTEXT

| □ | | A | B | C | D | E | F | G | H | I | J | K | L | M | N | O | P | Q | R | S | T | U | V | W | X | Y | Z | 0 | 1 | 2 | 3 | 4 | 5 | 6 | 7 | 8 | 9 |  | , | . | ( | ) | + | - | * | / | ^ | < | = | > | % | € | £ | $ |
|---|---|---|---|---|---|---|---|---|---|---|---|---|---|---|---|---|---|---|---|---|---|---|---|---|---|---|---|---|---|---|---|---|---|---|---|---|---|---|---|---|---|---|---|---|---|---|---|---|---|---|---|---|---|---|
| ranks → ↓ letters | ↓→ | 1 | 2 | 3 | 4 | 5 | 6 | 7 | 8 | 9 | 10 | 11 | 12 | 13 | 14 | 15 | 16 | 17 | 18 | 19 | 20 | 21 | 22 | 23 | 24 | 25 | 26 | 27 | 28 | 29 | 30 | 31 | 32 | 33 | 34 | 35 | 36 | 37 | 38 | 39 | 40 | 41 | 42 | 43 | 44 | 45 | 46 | 47 | 48 | 49 | 50 | 51 | 52 | 53 |
| A | 1 | A | B | C | D | E | F | G | H | I | J | K | L | M | N | O | P | Q | R | S | T | U | V | W | X | Y | Z | 0 | 1 | 2 | 3 | 4 | 5 | 6 | 7 | 8 | 9 |  | , | . | ( | ) | + | - | * | / | ^ | < | = | > | % | € | £ | $ |
| B | 2 | B | C | D | E | F | G | H | I | J | K | L | M | N | O | P | Q | R | S | T | U | V | W | X | Y | Z | 0 | 1 | 2 | 3 | 4 | 5 | 6 | 7 | 8 | 9 |  | , | . | ( | ) | + | - | * | / | ^ | < | = | > | % | € | £ | $ | A |
| C | 3 | C | D | E | F | G | H | I | J | K | L | M | N | O | P | Q | R | S | T | U | V | W | X | Y | Z | 0 | 1 | 2 | 3 | 4 | 5 | 6 | 7 | 8 | 9 |  | , | . | ( | ) | + | - | * | / | ^ | < | = | > | % | € | £ | $ | A | B |
| D | 4 | D | E | F | G | H | I | J | K | L | M | N | O | P | Q | R | S | T | U | V | W | X | Y | Z | 0 | 1 | 2 | 3 | 4 | 5 | 6 | 7 | 8 | 9 |  | , | . | ( | ) | + | - | * | / | ^ | < | = | > | % | € | £ | $ | A | B | C |
| E | 5 | E | F | G | H | I | J | K | L | M | N | O | P | Q | R | S | T | U | V | W | X | Y | Z | 0 | 1 | 2 | 3 | 4 | 5 | 6 | 7 | 8 | 9 |  | , | . | ( | ) | + | - | * | / | ^ | < | = | > | % | € | £ | $ | A | B | C | D |
| F | 6 | F | G | H | I | J | K | L | M | N | O | P | Q | R | S | T | U | V | W | X | Y | Z | 0 | 1 | 2 | 3 | 4 | 5 | 6 | 7 | 8 | 9 |  | , | . | ( | ) | + | - | * | / | ^ | < | = | > | % | € | £ | $ | A | B | C | D | E |
| G | 7 | G | H | I | J | K | L | M | N | O | P | Q | R | S | T | U | V | W | X | Y | Z | 0 | 1 | 2 | 3 | 4 | 5 | 6 | 7 | 8 | 9 |  | , | . | ( | ) | + | - | * | / | ^ | < | = | > | % | € | £ | $ | A | B | C | D | E | F |
| H | 8 | H | I | J | K | L | M | N | O | P | Q | R | S | T | U | V | W | X | Y | Z | 0 | 1 | 2 | 3 | 4 | 5 | 6 | 7 | 8 | 9 |  | , | . | ( | ) | + | - | * | / | ^ | < | = | > | % | € | £ | $ | A | B | C | D | E | F | G |
| I | 9 | I | J | K | L | M | N | O | P | Q | R | S | T | U | V | W | X | Y | Z | 0 | 1 | 2 | 3 | 4 | 5 | 6 | 7 | 8 | 9 |  | , | . | ( | ) | + | - | * | / | ^ | < | = | > | % | € | £ | $ | A | B | C | D | E | F | G | H |
| J | 10 | J | K | L | M | N | O | P | Q | R | S | T | U | V | W | X | Y | Z | 0 | 1 | 2 | 3 | 4 | 5 | 6 | 7 | 8 | 9 |  | , | . | ( | ) | + | - | * | / | ^ | < | = | > | % | € | £ | $ | A | B | C | D | E | F | G | H | I |
| K | 11 | K | L | M | N | O | P | Q | R | S | T | U | V | W | X | Y | Z | 0 | 1 | 2 | 3 | 4 | 5 | 6 | 7 | 8 | 9 |  | , | . | ( | ) | + | - | * | / | ^ | < | = | > | % | € | £ | $ | A | B | C | D | E | F | G | H | I | J |
| L | 12 | L | M | N | O | P | Q | R | S | T | U | V | W | X | Y | Z | 0 | 1 | 2 | 3 | 4 | 5 | 6 | 7 | 8 | 9 |  | , | . | ( | ) | + | - | * | / | ^ | < | = | > | % | € | £ | $ | A | B | C | D | E | F | G | H | I | J | K |
| M | 13 | M | N | O | P | Q | R | S | T | U | V | W | X | Y | Z | 0 | 1 | 2 | 3 | 4 | 5 | 6 | 7 | 8 | 9 |  | , | . | ( | ) | + | - | * | / | ^ | < | = | > | % | € | £ | $ | A | B | C | D | E | F | G | H | I | J | K | L |
| N | 14 | N | O | P | Q | R | S | T | U | V | W | X | Y | Z | 0 | 1 | 2 | 3 | 4 | 5 | 6 | 7 | 8 | 9 |  | , | . | ( | ) | + | - | * | / | ^ | < | = | > | % | € | £ | $ | A | B | C | D | E | F | G | H | I | J | K | L | M |
| O | 15 | O | P | Q | R | S | T | U | V | W | X | Y | Z | 0 | 1 | 2 | 3 | 4 | 5 | 6 | 7 | 8 | 9 |  | , | . | ( | ) | + | - | * | / | ^ | < | = | > | % | € | £ | $ | A | B | C | D | E | F | G | H | I | J | K | L | M | N |
| P | 16 | P | Q | R | S | T | U | V | W | X | Y | Z | 0 | 1 | 2 | 3 | 4 | 5 | 6 | 7 | 8 | 9 |  | , | . | ( | ) | + | - | * | / | ^ | < | = | > | % | € | £ | $ | A | B | C | D | E | F | G | H | I | J | K | L | M | N | O |
| Q | 17 | Q | R | S | T | U | V | W | X | Y | Z | 0 | 1 | 2 | 3 | 4 | 5 | 6 | 7 | 8 | 9 |  | , | . | ( | ) | + | - | * | / | ^ | < | = | > | % | € | £ | $ | A | B | C | D | E | F | G | H | I | J | K | L | M | N | O | P |
| R | 18 | R | S | T | U | V | W | X | Y | Z | 0 | 1 | 2 | 3 | 4 | 5 | 6 | 7 | 8 | 9 |  | , | . | ( | ) | + | - | * | / | ^ | < | = | > | % | € | £ | $ | A | B | C | D | E | F | G | H | I | J | K | L | M | N | O | P | Q |
| S | 19 | S | T | U | V | W | X | Y | Z | 0 | 1 | 2 | 3 | 4 | 5 | 6 | 7 | 8 | 9 |  | , | . | ( | ) | + | - | * | / | ^ | < | = | > | % | € | £ | $ | A | B | C | D | E | F | G | H | I | J | K | L | M | N | O | P | Q | R |
| T | 20 | T | U | V | W | X | Y | Z | 0 | 1 | 2 | 3 | 4 | 5 | 6 | 7 | 8 | 9 |  | , | . | ( | ) | + | - | * | / | ^ | < | = | > | % | € | £ | $ | A | B | C | D | E | F | G | H | I | J | K | L | M | N | O | P | Q | R | S |
| U | 21 | U | V | W | X | Y | Z | 0 | 1 | 2 | 3 | 4 | 5 | 6 | 7 | 8 | 9 |  | , | . | ( | ) | + | - | * | / | ^ | < | = | > | % | € | £ | $ | A | B | C | D | E | F | G | H | I | J | K | L | M | N | O | P | Q | R | S | T |
| V | 22 | V | W | X | Y | Z | 0 | 1 | 2 | 3 | 4 | 5 | 6 | 7 | 8 | 9 |  | , | . | ( | ) | + | - | * | / | ^ | < | = | > | % | € | £ | $ | A | B | C | D | E | F | G | H | I | J | K | L | M | N | O | P | Q | R | S | T | U |
| W | 23 | W | X | Y | Z | 0 | 1 | 2 | 3 | 4 | 5 | 6 | 7 | 8 | 9 |  | , | . | ( | ) | + | - | * | / | ^ | < | = | > | % | € | £ | $ | A | B | C | D | E | F | G | H | I | J | K | L | M | N | O | P | Q | R | S | T | U | V |
| X | 24 | X | Y | Z | 0 | 1 | 2 | 3 | 4 | 5 | 6 | 7 | 8 | 9 |  | , | . | ( | ) | + | - | * | / | ^ | < | = | > | % | € | £ | $ | A | B | C | D | E | F | G | H | I | J | K | L | M | N | O | P | Q | R | S | T | U | V | W |
| Y | 25 | Y | Z | 0 | 1 | 2 | 3 | 4 | 5 | 6 | 7 | 8 | 9 |  | , | . | ( | ) | + | - | * | / | ^ | < | = | > | % | € | £ | $ | A | B | C | D | E | F | G | H | I | J | K | L | M | N | O | P | Q | R | S | T | U | V | W | X |
| Z | 26 | Z | 0 | 1 | 2 | 3 | 4 | 5 | 6 | 7 | 8 | 9 |  | , | . | ( | ) | + | - | * | / | ^ | < | = | > | % | € | £ | $ | A | B | C | D | E | F | G | H | I | J | K | L | M | N | O | P | Q | R | S | T | U | V | W | X | Y |
| 0 | 27 | 0 | 1 | 2 | 3 | 4 | 5 | 6 | 7 | 8 | 9 |  | , | . | ( | ) | + | - | * | / | ^ | < | = | > | % | € | £ | $ | A | B | C | D | E | F | G | H | I | J | K | L | M | N | O | P | Q | R | S | T | U | V | W | X | Y | Z |
| 1 | 28 | 1 | 2 | 3 | 4 | 5 | 6 | 7 | 8 | 9 |  | , | . | ( | ) | + | - | * | / | ^ | < | = | > | % | € | £ | $ | A | B | C | D | E | F | G | H | I | J | K | L | M | N | O | P | Q | R | S | T | U | V | W | X | Y | Z | 0 |
| 2 | 29 | 2 | 3 | 4 | 5 | 6 | 7 | 8 | 9 |  | , | . | ( | ) | + | - | * | / | ^ | < | = | > | % | € | £ | $ | A | B | C | D | E | F | G | H | I | J | K | L | M | N | O | P | Q | R | S | T | U | V | W | X | Y | Z | 0 | 1 |
| 3 | 30 | 3 | 4 | 5 | 6 | 7 | 8 | 9 |  | , | . | ( | ) | + | - | * | / | ^ | < | = | > | % | € | £ | $ | A | B | C | D | E | F | G | H | I | J | K | L | M | N | O | P | Q | R | S | T | U | V | W | X | Y | Z | 0 | 1 | 2 |
| 4 | 31 | 4 | 5 | 6 | 7 | 8 | 9 |  | , | . | ( | ) | + | - | * | / | ^ | < | = | > | % | € | £ | $ | A | B | C | D | E | F | G | H | I | J | K | L | M | N | O | P | Q | R | S | T | U | V | W | X | Y | Z | 0 | 1 | 2 | 3 |
| 5 | 32 | 5 | 6 | 7 | 8 | 9 |  | , | . | ( | ) | + | - | * | / | ^ | < | = | > | % | € | £ | $ | A | B | C | D | E | F | G | H | I | J | K | L | M | N | O | P | Q | R | S | T | U | V | W | X | Y | Z | 0 | 1 | 2 | 3 | 4 |
| 6 | 33 | 6 | 7 | 8 | 9 |  | , | . | ( | ) | + | - | * | / | ^ | < | = | > | % | € | £ | $ | A | B | C | D | E | F | G | H | I | J | K | L | M | N | O | P | Q | R | S | T | U | V | W | X | Y | Z | 0 | 1 | 2 | 3 | 4 | 5 |
| 7 | 34 | 7 | 8 | 9 |  | , | . | ( | ) | + | - | * | / | ^ | < | = | > | % | € | £ | $ | A | B | C | D | E | F | G | H | I | J | K | L | M | N | O | P | Q | R | S | T | U | V | W | X | Y | Z | 0 | 1 | 2 | 3 | 4 | 5 | 6 |
| 8 | 35 | 8 | 9 |  | , | . | ( | ) | + | - | * | / | ^ | < | = | > | % | € | £ | $ | A | B | C | D | E | F | G | H | I | J | K | L | M | N | O | P | Q | R | S | T | U | V | W | X | Y | Z | 0 | 1 | 2 | 3 | 4 | 5 | 6 | 7 |
| 9 | 36 | 9 |  | , | . | ( | ) | + | - | * | / | ^ | < | = | > | % | € | £ | $ | A | B | C | D | E | F | G | H | I | J | K | L | M | N | O | P | Q | R | S | T | U | V | W | X | Y | Z | 0 | 1 | 2 | 3 | 4 | 5 | 6 | 7 | 8 |
|  | 37 |  | , | . | ( | ) | + | - | * | / | ^ | < | = | > | % | € | £ | $ | A | B | C | D | E | F | G | H | I | J | K | L | M | N | O | P | Q | R | S | T | U | V | W | X | Y | Z | 0 | 1 | 2 | 3 | 4 | 5 | 6 | 7 | 8 | 9 |
| , | 38 | , | . | ( | ) | + | - | * | / | ^ | < | = | > | % | € | £ | $ | A | B | C | D | E | F | G | H | I | J | K | L | M | N | O | P | Q | R | S | T | U | V | W | X | Y | Z | 0 | 1 | 2 | 3 | 4 | 5 | 6 | 7 | 8 | 9 |  |
| . | 39 | . | ( | ) | + | - | * | / | ^ | < | = | > | % | € | £ | $ | A | B | C | D | E | F | G | H | I | J | K | L | M | N | O | P | Q | R | S | T | U | V | W | X | Y | Z | 0 | 1 | 2 | 3 | 4 | 5 | 6 | 7 | 8 | 9 |  | , |
| ( | 40 | ( | ) | + | - | * | / | ^ | < | = | > | % | € | £ | $ | A | B | C | D | E | F | G | H | I | J | K | L | M | N | O | P | Q | R | S | T | U | V | W | X | Y | Z | 0 | 1 | 2 | 3 | 4 | 5 | 6 | 7 | 8 | 9 |  | , | . |
| ) | 41 | ) | + | - | * | / | ^ | < | = | > | % | € | £ | $ | A | B | C | D | E | F | G | H | I | J | K | L | M | N | O | P | Q | R | S | T | U | V | W | X | Y | Z | 0 | 1 | 2 | 3 | 4 | 5 | 6 | 7 | 8 | 9 |  | , | . | ( |
| + | 42 | + | - | * | / | ^ | < | = | > | % | € | £ | $ | A | B | C | D | E | F | G | H | I | J | K | L | M | N | O | P | Q | R | S | T | U | V | W | X | Y | Z | 0 | 1 | 2 | 3 | 4 | 5 | 6 | 7 | 8 | 9 |  | , | . | ( | ) |
| - | 43 | - | * | / | ^ | < | = | > | % | € | £ | $ | A | B | C | D | E | F | G | H | I | J | K | L | M | N | O | P | Q | R | S | T | U | V | W | X | Y | Z | 0 | 1 | 2 | 3 | 4 | 5 | 6 | 7 | 8 | 9 |  | , | . | ( | ) | + |
| * | 44 | * | / | ^ | < | = | > | % | € | £ | $ | A | B | C | D | E | F | G | H | I | J | K | L | M | N | O | P | Q | R | S | T | U | V | W | X | Y | Z | 0 | 1 | 2 | 3 | 4 | 5 | 6 | 7 | 8 | 9 |  | , | . | ( | ) | + | - |
| / | 45 | / | ^ | < | = | > | % | € | £ | $ | A | B | C | D | E | F | G | H | I | J | K | L | M | N | O | P | Q | R | S | T | U | V | W | X | Y | Z | 0 | 1 | 2 | 3 | 4 | 5 | 6 | 7 | 8 | 9 |  | , | . | ( | ) | + | - | * |
| ^ | 46 | ^ | < | = | > | % | € | £ | $ | A | B | C | D | E | F | G | H | I | J | K | L | M | N | O | P | Q | R | S | T | U | V | W | X | Y | Z | 0 | 1 | 2 | 3 | 4 | 5 | 6 | 7 | 8 | 9 |  | , | . | ( | ) | + | - | * | / |
| < | 47 | < | = | > | % | € | £ | $ | A | B | C | D | E | F | G | H | I | J | K | L | M | N | O | P | Q | R | S | T | U | V | W | X | Y | Z | 0 | 1 | 2 | 3 | 4 | 5 | 6 | 7 | 8 | 9 |  | , | . | ( | ) | + | - | * | / | ^ |
| = | 48 | = | > | % | € | £ | $ | A | B | C | D | E | F | G | H | I | J | K | L | M | N | O | P | Q | R | S | T | U | V | W | X | Y | Z | 0 | 1 | 2 | 3 | 4 | 5 | 6 | 7 | 8 | 9 |  | , | . | ( | ) | + | - | * | / | ^ | < |
| > | 49 | > | % | € | £ | $ | A | B | C | D | E | F | G | H | I | J | K | L | M | N | O | P | Q | R | S | T | U | V | W | X | Y | Z | 0 | 1 | 2 | 3 | 4 | 5 | 6 | 7 | 8 | 9 |  | , | . | ( | ) | + | - | * | / | ^ | < | = |
| % | 50 | % | € | £ | $ | A | B | C | D | E | F | G | H | I | J | K | L | M | N | O | P | Q | R | S | T | U | V | W | X | Y | Z | 0 | 1 | 2 | 3 | 4 | 5 | 6 | 7 | 8 | 9 |  | , | . | ( | ) | + | - | * | / | ^ | < | = | > |
| € | 51 | € | £ | $ | A | B | C | D | E | F | G | H | I | J | K | L | M | N | O | P | Q | R | S | T | U | V | W | X | Y | Z | 0 | 1 | 2 | 3 | 4 | 5 | 6 | 7 | 8 | 9 |  | , | . | ( | ) | + | - | * | / | ^ | < | = | > | % |
| £ | 52 | £ | $ | A | B | C | D | E | F | G | H | I | J | K | L | M | N | O | P | Q | R | S | T | U | V | W | X | Y | Z | 0 | 1 | 2 | 3 | 4 | 5 | 6 | 7 | 8 | 9 |  | , | . | ( | ) | + | - | * | / | ^ | < | = | > | % | € |
| $ | 53 | $ | A | B | C | D | E | F | G | H | I | J | K | L | M | N | O | P | Q | R | S | T | U | V | W | X | Y | Z | 0 | 1 | 2 | 3 | 4 | 5 | 6 | 7 | 8 | 9 |  | , | . | ( | ) | + | - | * | / | ^ | < | = | > | % | € | £ |

Alphabet : ABCDEFGHIJKLMNOPQRSTUVWXYZ0123456789 ,.:’'"()+-*/^<=>&@%€£$

Plaintext* : SPIRALE is a one-time pad cryptosystem designed in 2015/05 to replace SOLITAIRE when one has no cards. It is based in parts on the generalized Fibonacci sequence xn = xn-49 * xn-24. SPIRALE is free and open source.

* Python program : as there is no symbol for end of line or carrier return in this alphabet, the plaintext must be in a continuous line.

Keys : NVIKKIH CTSQEOU DNGDKSZ EAIWDSH

Permuted Alphabet 1 : +XOD='7S&"3QG€,J£’0NC:H<6RA-V9U%5M@Y1E8B KW$2)^L4I*(PZ/>T.F

Permuted Alphabet 2 : € RA&.PM/1HC"SN+W%^3£=X*YO)JF2(LD5,7’>G<I$'6VQ-8:0UE9BK@TZ4

Ciphertext* : DY2R"S7->EQFS@MT&1T@X%*"AHE:9QR@F@@TDT€£0DECK$N"'NO8>P:E£*H0 X’C 4FI7YC693&=&-
£.K^.A72J*.O’RG)(S820XX<YX/U(€PQWCN/,GL(5XE75"PNEP,7ILC£U:- ,-D5R9,€U5+V£CE"RQ"PW2/ ",V"NJ+-
'>H5"7OUK:=@NH/6$*6RXFLT^64U./@L9S$'=(G:9<U*

*Python program : ciphertext is thus also a continuous line.

```
PythonWin 3.4.1 (default, Aug  7 2014, 13:13:27) [MSC v.1600 32 bit (Intel)] on win32.
Portions Copyright 1994-2008 Mark Hammond - see 'Help/About PythonWin' for further copyright information.
>>> SPIRALE 2.1.3:
Encryption process from keys (NVIKKIH|CTSQEOU|DNGDKSZ|EAIWDSH) and 59-alphabet
[ABCDEFGHIJKLMNOPQRSTUVWXYZ0123456789 ,.:’'"()+-*/^<=>&@%€£$]

1. Generate ciphering table from key 1 and key 2:
Permutation of alphabet according to following ranks : [14, 22, 9, 11, 11, 9, 8]  for
NVIKKIH:
Permuted alphabet (59 characters) = +XOD='7S&"3QG€,J£’0NC:H<6RA-V9U%5M@Y1E8B
KW$2)^L4I*(PZ/>T.F
The permuted alphabet was written to file PermAlph_NVIKKIH.txt
Permutation of alphabet according to following ranks : [3, 20, 19, 17, 5, 15, 21]  for
CTSQEOU:
Permuted alphabet (59 characters) = €
RA&.PM/1HC"SN+W%^3£=X*YO)JF2(LD5,7’>G<I$'6VQ-8:0UE9BK@TZ4
The permuted alphabet was written to file PermAlph_CTSQEOU.txt

2. Generate long key from key 3 and key 4:
Product matrix by key 3 and key 4:
['@', 'G', '(', 'T', '9', 'Q', 'N']
['L', 'W', 'A', '9', '=', '6', '3']
['E', 'P', '>', '2', ')', 'Z', 'W']
['@', 'G', '(', 'T', '9', 'Q', 'N']
['7', ')', 'W', '£', 'O', '@', '=']
['$', 'K', '*', 'X', ':', 'U', 'R']
['+', '€', '8', 'K', '0', 'H', 'E']
Generated long key (49 characters) = @LGEW(@PAT7G>99$)(2=Q+KWT)6N€*£9Z38XOQWK:@N0U=HRE

3. Generate keystream:
Keystream (214 characters) =
@LGEW(@PAT7G>99$)(2=Q+KWT)6N€*£9Z38XOQWK:@N0U=HREB50,GM1<+)2,'+Y6'G8@6K(8<TO&=£ZJJ
I9)0HZENHOO$>NR £W965E<B0,’@K7QTWEAGTG’YRSB*^+@J4YG'N£:89OF’’8%^"3
>"*7<.">A/C@$^IGD/)39*)IRDK9OUBNPN4MD€K’5Z(£YK9<"2E(5O=XC’&1H('0
The keystream sequence of characters was written to file keystream.txt.

4. Encryption with OTP:
The generated ciphertext was written to 'ciphertext.txt'.
Plaintext:  SPIRALE IS A ONE-TIME PAD CRYPTOSYSTEM DESIGNED IN 2015/05 TO REPLACE
SOLITAIRE WHEN ONE HAS NO CARDS. IT IS BASED IN PARTS ON THE GENERALIZED FIBONACCI
SEQUENCE XN = XN-49 * XN-24. SPIRALE IS FREE AND OPEN SOURCE.
Ciphertext: DY2R"S7->EQFS@MT&1T@X%*"AHE:9QR@F@@TDT€£0DECK$N"'NO8>P:E£*H0 X’C
4FI7YC693&=&-£.K^.A72J*.O’RG)(S820XX<YX/U(€PQWCN/,GL(5XE75"PNEP,7ILC£U:-
,-D5R9,€U5+V£CE"RQ"PW2/ ",V"NJ+-'>H5"7OUK:=@NH/6$*6RXFLT^64U./@L9S$'=(G:9<U*
```